\documentclass[a4paper,12pt,notitlepage]{article}
\usepackage{latexsym}
\usepackage{amssymb}
\usepackage{eufrak}
\usepackage[dvips]{graphics}

\title{Tri--hamiltonian vector fields, spectral
curves and separation coordinates}
\author{\textbf{L. Degiovanni} and \textbf{G. Magnano}\\
Dipartimento di Matematica, Universit\`a di
Torino\\
Via Carlo Alberto 10, I--10131 Torino, Italy}
\date{ }

\newcommand{\BD}{\begin{displaymath}} 
\newcommand{\ED}{\end{displaymath}}
\newcommand{\BE}{\begin{equation}}
\newcommand{\EE}{\end{equation}}

\newcommand{\D}{\mathrm{d}}
\newcommand{\identity}{\mathrm{1\kern-3pt{I}}}

\newcommand{\half}{\frac{1}{2}}
\newcommand{\tr}{Tr}
\newcommand{\tder}[1]{\frac{d#1}{dt}}
\newcommand{\Pp}{P_0}
\newcommand{\Q}{P_1}
\newcommand{\R}{P_2}
\newcommand{\Nq}{N_1}
\newcommand{\Nr}{N_2}
\newcommand{\Pn}{P^{(n)}}
\newcommand{\lcdot}{\,\cdot\ }

\newcommand{\Doppio}[1]{\mathbb{#1}}

\newcommand{\Gotico}[1]{\mathfrak{#1}}
\newcommand{\Reali}{\Doppio{R}}
\newcommand{\Complessi}{\Doppio{C}}

\newcommand{\so}{\Gotico{so}}
\newcommand{\gl}{\Gotico{gl}}

\newcommand{\Scal}[2]{(#1,#2)}

\newcommand{\Dual}[2]{\langle #1,#2 \rangle}

\newcommand{\Pois}[2]{\{#1,#2\}}

\newcommand{\Lie}[2]{\lbrack #1,#2 \rbrack}

\newcommand{\Triplo}[3]{#1#2#3 - #3#2#1}

\newcommand{\Schouten}[2]{{\lbrack #1,#2 \rbrack}_\mathrm{S}}

\newcommand{\auno}{\alpha}
\newcommand{\adue}{\beta}
\newcommand{\atre}{\gamma}

\newcounter{def}

\newenvironment{Def}%
{\begin{quotation}\noindent\textbf{Defininition\addtocounter{def}{1} 
\thesection.\thedef}:}%
{\end{quotation}}

\newcounter{teo}

\newenvironment{Prop}[1]%
{\medskip\goodbreak\noindent\textbf{Proposition\addtocounter{teo}{1}
\thesection.\theteo}#1:\slshape}%
{\upshape}

\newenvironment{Lemma}%
{\medskip\goodbreak\noindent\textbf{Lemma\addtocounter{teo}{1} 
\thesection.\theteo}:\slshape}%
{\upshape}

\newenvironment{Dim}%
{\par\medskip\upshape\noindent\textbf{Proof}:}%
{\hfill$\Box$\bigskip}

\newsavebox{\Pdir}
\savebox{\Pdir}(15,10){
\put(0,0){\vector(3,2){15}}
\put(0,6){\tiny $\Pp$}
}

\newsavebox{\Pinv}
\savebox{\Pinv}(15,10){
\put(15,10){\vector(-3,-2){15}}
\put(5,7){\tiny $\Pp$}
}

\newsavebox{\Qdir}
\savebox{\Qdir}(15,10){
\put(15,0){\vector(-3,2){15}}
\put(8,6){\tiny $\Q$}
}

\newsavebox{\Qinv}
\savebox{\Qinv}(15,10){
\put(0,10){\vector(3,-2){15}}
\put(6,7){\tiny $\Q$}
}

\newsavebox{\QLenD}
\savebox{\QLenD}(15,10){
\put(0,0){\vector(3,2){15}}
\put(0,6){\tiny $\Q$}
}

\newsavebox{\QLenS}
\savebox{\QLenS}(15,10){
\put(15,0){\vector(-3,2){15}}
\put(8,6){\tiny $\Q$}
}

\newsavebox{\Rdir}
\savebox{\Rdir}(15,18){
\put(7,18){\vector(0,-1){18}}
\put(8,8){\tiny $\R$}
}

\newsavebox{\Rinv}
\savebox{\Rinv}(15,18){
\put(7,0){\vector(0,1){18}}
\put(8,6){\tiny $\R$}
}

\newsavebox{\RLenD}
\savebox{\RLenD}(15,10){
\put(0,0){\vector(3,2){15}}
\put(0,6){\tiny $\R$}
}

\newsavebox{\RLenS}
\savebox{\RLenS}(15,10){
\put(15,0){\vector(-3,2){15}}
\put(8,6){\tiny $\R$}
}

\newsavebox{\Palt}
\savebox{\Palt}(30,10){
\put(15,0){\vector(-3,2){15}}
\put(8,6){\tiny $\Pp$}
}

\newsavebox{\Qalt}
\savebox{\Qalt}(15,10){
\put(0,0){\vector(3,2){15}}
\put(3,6){\tiny $\Q$}
}

\newsavebox{\Qqdir}
\savebox{\Qqdir}(15,10){
\put(15,0){\vector(-3,2){15}}
\put(8,6){\tiny $\Q$}
}

\newcommand{\Scatola}[3]{
\put(#1,#2){\put(7.5,0){\makebox(15,15){#3}}}
}

\newcommand{\Scatolona}[3]{
\put(#1,#2){\put(7.5,0){\makebox(25,25){#3}}}
}

\newcommand{\NucDir}[1]{%
\begin{picture}(45,43)
\put(0,0){\usebox{\Pdir}}
\put(15,25){\usebox{\Rdir}}
\put(30,0){\usebox{\Qdir}}
\Scatola{15}{10}{#1}
\end{picture}
}

\newcommand{\NucAlt}[1]{%
\begin{picture}(55,50)
\put(2,40){\usebox{\Qdir}}
\put(20,0){\usebox{\Rdir}}
\put(38,40){\usebox{\Pdir}}
\Scatolona{15}{16.5}{#1}
\end{picture}
}

\begin{document}

\maketitle

\begin{abstract}
We show that for a class of dynamical systems, Hamiltonian
with respect to three distinct Poisson brackets $(P_0,P_1,P_2)$,
separation coordinates are provided by the common roots of a set of
bivariate polynomials.
These polynomials, which generalise those considered by E.~Sklyanin
in his algebro--geometric approach, are obtained from the knowledge
of:
(i) a common Casimir function for the two Poisson
pencils $(P_1-\lambda P_0)$ and $(P_2-\mu P_0)$;
(ii) a suitable set of vector fields, preserving $\Pp$ but transversal to
its symplectic leaves.
The frameworks is applied to Lax equations with spectral parameter, for
which not only it establishes a theoretical link between the
separation techniques of Sklyanin and of Magri, but also provides a more
efficient ``inverse'' procedure to obtain separation variables, not
involving the extraction of roots.
\end{abstract}

\section{Introduction}

The relationship between the Liouville integrability of a Hamiltonian
system and the existence
of a second conserved Poisson bracket (or "hamiltonian structure") in its
phase space, first discovered
by Magri \cite{Magri1}, has been thoroughly investigated in the past years.
Bihamiltonian structures
underlying all classical examples of integrable systems (both finite and
infinite-dimensional) have
been described by several authors, and almost all the relevant properties
connected to integrability
have been reinterpreted in terms of the geometry of bihamiltonian manifolds
and vector fields.
Recently, the classical problem of characterizing separable hamiltonians
(i.e.~those for which the
Hamilton--Jacobi equation can be solved by separation of variables in a
suitable system of canonical
coordinates) has been also translated in the language of bihamiltonian
geometry \cite{Magri2}\cite{Magri2b}.

A question which has not yet received a complete answer concerns the link
between the bihamiltonian
framework and the algebro-geometric methods of solution based on the
isospectrality property of Lax equations \cite{Grif} \cite{Van}. Although it is
possible to introduce
bihamiltonian structures which naturally lead to Lax equations with a
spectral parameter
\cite{MagnMagri} \cite{ReySem1}, the role of the characteristic 
equation for the Lax
operator (the
"spectral curve" of the algebro--geometric approach) has not been clarified
so far in the
bihamiltonian perspective.

The present work adds new elements in view of a connection between
multihamiltonian structures,
existence of separation coordinates and spectral curves, starting from an
apparently marginal observation:
some well--known integrable systems allow two distinct
bihamiltonian descriptions, independently described by different authors
and apparently unrelated (in spite
of having one Poisson bracket in common). In this introductory section, we
shall recall some relevant facts using the
simplest example of such ``trihamiltonian systems", namely the generalized
Euler--Poinsot rigid body. To
motivate the reader to follow us through an exercise which could seem of
little practical interest, let
us anticipate that the interplay of the three Poisson
structures leads to a new role played by
the characteristic determinant of the Lax matrix, and this fact may
eventually clarify the connection between Sklyanin's algebro--geometrical
construction of separation variables
\cite{Skly} and the bihamiltonian method recently
proposed in \cite{Magri2}\cite{Magri2b}.

Indeed, the occurrence of more than two Poisson brackets on the same
manifold is not new nor
surprising by itself, and in some cases it is even a structural property,
as for the so--called
"Lie--Poisson pencils" described in \cite{MagnMagri}; in the sequel, we
discuss the difference
between such known cases of multihamiltonian structures and the
trihamiltonian structure that we are presently considering.

\bigskip
The simplest (nontrivial) example of Lax equation with spectral parameter
is provided by the
dynamics of a rigid body about a
fixed point, in the absence of external forces (Poinsot rigid body). In the
body reference
frame, the motion is described by the Euler--Poisson--Lax equation
\BE
\tder{M} = [M,\Omega], \label{EP1}
\EE
where $M$ and $\Omega$ are the skew--symmetric $3\times3$ matrices 
representing the angular momentum
and the angular velocity, respectively, in the body reference frame. 
A straightforward
consequence of (\ref{EP1}) is that the trace of any power of the 
matrix $M$ is a
constant of motion: $\tder{}\tr(M^k) = 0$.
Generalizing the system to $M\in\so(r)$, one obtains in this way at
most $\frac{r}{2}$ independent constants of motion (if $r$ is even, or
$\frac{r-1}{2}$ for $r$
odd), which for $r>3$ would not be enough to meet Liouville's integrability
condition.

Assuming
$(I_1,I_2,I_3)$ to be the eigenvalues of the inertia tensor, one can 
introduce the
diagonal matrix with
diagonal elements
\mbox{$(\frac{-I_1+I_2+I_3}{2},\frac{I_1-I_2+I_3}{2},\frac{I_1+I_2-I_3}{2})$};
the linear
relation between
$M$ and
$\Omega$ can then be written in the following form:
\BE
M=J \Omega+\Omega J ;\label{EP2}
\EE
Manakov \cite{Mana} has observed that the Euler--Poisson equation
(\ref{EP1}) and the inertia map
(\ref{EP2}) can be put together into a single Lax equation for a new Lax
pair dependent on a
formal parameter $\lambda$,
\BE
\tder{(M+\lambda J^2)} =[M+\lambda J^2, \Omega+\lambda J].
\label{EPM}
\EE
The insertion of the parameter $\lambda$ into the Lax equation leads
to a wider
number of constants of motion, the {\it Manakov integrals\/}. We denote them by
$f_i^k$, according to
the following convention:
\BE
f^{(k)}_\lambda=\frac{1}{k}J^{2k}\lambda^{k}+\sum_{i=1}^k f^k_i\lambda^{k-i}.
\EE
For $M\in\so(r)$, the functions $f_i^k(M)$ vanish identically for
$i$ odd; the odd Manakov functions are however relevant for the
``generalized Euler--Poinsot system'', with $M\in\gl(r)$, that we shall
consider in the sequel.

As is well known, the equation (\ref{EP1}) is Hamiltonian with respect to
the {\it Lie-Poisson bracket\/}, defined on $\so(r)$ through the
ad--invariant scalar product
\BE
(A,B) = \tr(A\cdot B).\label{metrica}
\EE
More precisely, given any function $f:\so(r)\rightarrow\Reali$, one defines
the gradient
at a point $M$ as the matrix $\nabla f\in\so(r)$ such that $\dot f =
\langle df,\dot M\rangle =
(\dot M,\nabla f)$; then, for any pair of functions,
\BE
\Pois{f}{g}_{_{LP}} = (M,[\nabla f,\nabla g]) \label{LP}
\EE
is a Poisson bracket \cite{ReySem2}. The Lie--Poisson bracket (\ref{LP}) is
degenerate: an
ad--invariant function $f(M)$ is in involution with any other
function, i.e.~is a \emph{Casimir function} for the bracket 
(\ref{LP}). The Casimir functions,
which include the traces of the powers of $M$, are automatically 
constants of motion, but
they are irrelevant as far as the Liouville integrability of the 
system (with the
Lie--Poisson bracket) is concerned. Therefore, the
integrability of the system actually relies on the existence of the other
Manakov first integrals (among
which one can find enough independent functions).

In 1996, Morosi and Pizzocchero \cite{MorPizz} introduced a second Poisson
bracket on $\so(r)$, defined
as follows: let $A\in \gl(r)$ a fixed matrix (for the Euler--Poinsot case,
$A\equiv J^2$; notice that
$A$ needs not to belong to $\so(r)$). With the same definition of scalar
product and gradient as
above, one sets
\BE
\Pois{f}{g}_{_{MP}} = (M,\nabla f\cdot A\cdot \nabla g-\nabla g\cdot A\cdot
\nabla f). \label{MP}
\EE
One can check that the vector field generated by the Manakov functions through
the Poisson structure (\ref{MP}) are exactly the same as those generated
through the Lie--Poisson
structure (\ref{LP}), up to a rearrangement in the correspondence between
hamiltonians and vector
fields. For instance, the physical hamiltonian generating the
Euler--Poinsot dynamics through the
Lie--Poisson bracket is $h_1=\half\tr(\Omega
M)=\tr(\Omega^2 J)$, while the hamiltonian of the
same vector field through the Morosi--Pizzocchero bracket is
$h_2=-\half\tr(\Omega J^{-1} M J^{-1})$. To simplify the notation, let us
denote by
$\Q$ and $\R$ the Poisson tensors associated respectively to the brackets
(\ref{LP}) and (\ref{MP}):
\BE
\Pois{f}{g}_{_{LP}}=\langle df,\Q dg\rangle,\qquad
\Pois{f}{g}_{_{MP}}=\langle df,\R dg\rangle.
\label{pbra}
\EE
Denoting by $X_1$ the vector field over
$\so(r)$ corresponding to equation (\ref{EP1}), the relation
$\Q dh_1 = \R dh_2$ is depicted by the diagram
\BE\label{pezzo}
\begin{picture}(80,50)
\Scatola{0}{0}{$h_1$}
\put(15,15){\usebox{\QLenD}}
\Scatola{30}{25}{$X_1$}
\put(45,15){\usebox{\RLenS}}
\Scatola{60}{0}{$h_2$}
\end{picture}
\EE
where $h\stackrel{P}{\rightarrow} X$ is an abbreviation for
$dh\stackrel{P}{\longmapsto}
X$, a convention that we shall use in analogous diagrams throughout this
article. The diagram
(\ref{pezzo}) is nothing but the elementary block of the {\it Lenard--Magri
recursion\/} generating a
whole family of quadratic first integrals $h_i$ (known as
\emph{Mi\v{s}\v{c}enko functions}), and the
corresponding symmetry vector fields
$X_i$:
\BE\label{lenard}
\begin{picture}(300,50)
\multiput(15,15)(60,0){5}{%
\put(0,0){\usebox{\QLenD}}
}
\multiput(45,15)(60,0){4}{%
\put(0,0){\usebox{\RLenS}}
}
\Scatola{0}{0}{$h_1$}
\Scatola{30}{25}{$X_1$}
\Scatola{60}{0}{$h_2$}
\Scatola{90}{25}{$X_2$}
\Scatola{120}{0}{$h_3$}
\Scatola{150}{25}{$\cdots$}
\Scatola{180}{0}{$h_k$}
\Scatola{210}{25}{$X_k$}
\Scatola{240}{0}{$h_{k+1}$}
\Scatola{270}{25}{$\cdots$}
\end{picture}
\EE
The Manakov first integrals can be generated by the same recursion
procedure. Setting $A\equiv
J^2$, one has
\mbox{$\nabla f^{(k)}_\lambda = (M+\lambda A)^{k-1}$}, then
\BE
\Q df^k_i=[M,\nabla f^k_i]=A\nabla f^{k-1}_{i}M-M\nabla f^{k-1}_{i}A=\R
df^{k-1}_{i}
\EE
which correspond to Lenard--Magri diagrams starting with the $\Q$--Casimir
functions $f^k_k$:

\begin{picture}(200,80)
\multiput(65,25)(80,0){4}{%
\put(0,0){\usebox{\RLenD}}
}
\multiput(25,25)(80,0){4}{%
\put(0,0){\usebox{\QLenS}}
}
\Scatolona{0}{35}{$0$}
\Scatolona{40}{0}{$f_k^k$}
\Scatolona{80}{35}{$X_k^{k+1}$}
\Scatolona{120}{0}{$f_k^{k+1}$}
\Scatolona{160}{35}{$X_k^{k+2}$}
\Scatolona{200}{0}{$\cdots$}
\Scatolona{240}{35}{$X_k^{k+i}$}
\Scatolona{280}{0}{$f_k^{k+i}$}
\Scatolona{320}{35}{$\cdots$}
\end{picture}

Notice that all the functions iteratively generated by Lenard--Magri
recursion relations are
automatically in involution with respect to both Poisson tensors $\Q$ and
$\R$. The (elementary) proof of
this fact will be recalled in the next section. Thanks to this property of
bihamiltonian vector fields,
one does not need to prove separately the involutivity of the first
integrals of Manakov, and the complete
integrability of the generalized Euler--Poinsot system is simply assessed
by computing how many
independent first integrals can be found in this way.

All the statements above hold valid if one extends the equation (\ref{EP1})
to $M\in\gl(r)$. Both the
Lie--Poisson bracket and the Morosi--Pizzocchero bracket can be introduced
in $\gl(r)$ using the same
definitions (\ref{LP}) and (\ref{MP}). The
Morosi--Pizzocchero bracket is defined in terms of the matrix product (not
of the commutator) and
therefore is defined even more naturally on $\gl(r)$: it reduces on $\so(r)$ by
orthogonal projection with
respect to the scalar product (\ref{metrica}), provided the matrix $A$ is
symmetric. Thus, for the Lax matrix
$L(\lambda)=A\lambda+M$ the complete family
of Manakov constants of motion can be found by the recursion procedure,
which also ensures their mutual involutivity.
Whenever $A$ is symmetric (and positive), the dynamical system defined by
(\ref{EP1}) in $\gl(r)$ is
a proper extension of the original Euler--Poinsot system. The flows of the
original model
are those associated to the even Manakov functions (these flows are tangent
to $\so(r)$), while the
other flows of the enlarged system, generated by the odd Manakov functions, are
orthogonal to $\so(r)$.

In the larger phase space $\gl(r)$, however, one can obtain the full set of
first integrals
by \emph{another} Lenard--Magri recursion, relative to a \emph{different}
bihamiltonian pair. The new Poisson
bracket depends, as for (\ref{MP}), on the choice of the matrix $A$:
\BE
\Pois{f}{g}_{_{A}} = (A,[\nabla g,\nabla f]). \label{Q}
\EE
From now on, let us denote by $\Pp$ the Poisson tensor associated with this
bracket. The Manakov
functions are in bihamiltonian recursion also with respect to the pair
$(\Pp,\Q)$, but the sequences
are arranged in a different way:
\BE
\Q df^k_i=[M,\nabla f^k_i]=[\nabla f^{k}_{i+1},A]= \Pp df^k_{i+1}.
\EE
Thus, each integer power of $L(\lambda)$ corresponds to a single
finite Lenard--Magri sequence,
starting from a Casimir function for $\Pp$ and ending with a Casimir
function for $\Q$:
\BE\label{LMM}
\begin{picture}(420,70)
\multiput(65,25)(80,0){4}{%
\put(0,0){\usebox{\Pdir}}
}
\multiput(25,25)(80,0){4}{%
\put(0,0){\usebox{\Qdir}}
}
\Scatolona{0}{35}{$0$}
\Scatolona{40}{0}{$f_k^k$}
\Scatolona{80}{35}{$X_k^k$}
\Scatolona{120}{0}{$\cdots$}
\Scatolona{160}{35}{$X_2^k$}
\Scatolona{200}{0}{$f_2^k$}
\Scatolona{240}{35}{$X_1^k$}
\Scatolona{280}{0}{$f_1^k$}
\Scatolona{320}{35}{$0$}
\end{picture}
\EE
A disadvantage of the new bihamiltonian structure $(\Pp,\Q)$ is that it
\emph{cannot} be reduced (by restriction or by orthogonal projection)
to $\so(r)$.
On the other hand, $(\Pp,\Q)$ leads naturally to the Lax equation with
spectral parameter (\ref{EPM}),
which on the contrary is rather difficult to derive from the former pair
$(\Q,\R)$. To show this, we
need to reexpress the Lenard--Magri recursion relations (\ref{LMM}) in the
language of {\it Poisson
pencils}.

Given a pair of Poisson tensors $(P,Q)$ on a manifold
$\cal M$, assume that the $\lambda$--dependent bracket
\BE
\Pois{f}{g}_{_{P-\lambda
Q}}=\Pois{f}{g}_{_{P}}-\lambda\Pois{f}{g}_{_{Q}}=\langle
df,(P-\lambda Q) dg\rangle;\label{PP}
\EE
be a Poisson bracket, i.e.~fulfill the Jacobi identity for any
$\lambda$; in this case,
$P$ and $Q$ are said to be {\it compatible\/}; $(\mathcal{M}, P, Q)$
becomes a \emph{bihamiltonian manifold} (or
\emph{$PQ$--manifold}, following \cite{Magri3}), and one refers to the
bracket (\ref{PP}) as to its {\it
Poisson pencil}. It is immediate to see that, given a sequence of functions
$\lbrace
f_i\rbrace_{i=0,\ldots,N}$ such that
\begin{eqnarray}\label{cas1}
0 &=& P df_0 \nonumber\\
Q df_0 &=& P df_1 \nonumber\\[-6pt]
&\vdots& \\[-6pt]
Q df_k &=& P df_{k+1} \nonumber\\[-6pt]
&\vdots& \nonumber\\[-6pt]
Q df_N &=& 0 ,\nonumber
\end{eqnarray}
then the polynomial in $\lambda$ defined by
\BE
f_{\lambda}=f_0 + f_1\lambda + \cdots + f_N\lambda^N
\EE
is a Casimir function of the Poisson pencil, i.e.~for any $\lambda$
\BE
(P-\lambda Q)\ df_{\lambda}=0\label{cas}
\EE
(the differential of $f_{\lambda}$ is taken with respect to the coordinates
on $\cal M$, $\lambda$
being regarded as a parameter). Conversely, given a
$\lambda$--polynomial function fulfilling (\ref{cas}), its coefficients
obey the Magri--Lenard
recursion according to (\ref{cas1}) and generate a sequence of commuting
bihamiltonian
vector fields.

In the next section, we will recall the proof of the following relevant
property, that
we shall extensively use. {\em Let
$g_{\lambda}$ be a second Casimir function of the same Poisson pencil:
then, not only its
coefficients $g_k$ are in involution among themselves, but they also
Poisson--commute with all the
coefficients $f_k$ of the other Casimir function $f_{\lambda}$.}

Given a polynomial Casimir function $f_{\lambda}$, each bihamiltonian
vector field of the
associated Lenard--Magri hierarchy
$X_k = P df_k = Q df_{k-1}$ can be also represented by a {\it Hamilton
equation with spectral
parameter\/}. Having set $f_{\lambda}=\sum_{i=0}^{N}f_i\lambda^i$, consider
for each
positive integer $k<N$ the polynomial
\BE\label{hamspec1}
f^{(k)}_{\lambda}\equiv f_0\lambda^k + f_1\lambda^{k-1} + \cdots +
f_{k-1}\lambda + f_k;
\EE
taking into account (\ref{cas1}) it is easy to see that
\BE\label{hamspec2}
X_k = (P-\lambda Q)\ df^{(k)}_{\lambda}.\label{HSP}
\EE
This formula holds true for formal power series ($N=\infty)$; if the
Casimir function
$f_{\lambda}$ is instead expanded in Laurent series,
$f_{\lambda}=\sum_{i=0}^{\infty}f_i\lambda^{-i}$, then the polynomial
$f^{(k)}_{\lambda}$ is easily
obtained by multiplication by $\lambda^k$ and truncation to the nonnegative
powers:
\BE
f^{(k)}_{\lambda} = \left(\lambda^k f_{\lambda}\right)_+\ .
\EE

We are now ready to derive the Manakov equation (\ref{EPM}) as a Hamilton
equation with spectral
parameter for the Poisson pencil $(\Q-\lambda\Pp)$ on $\gl(r)$.

In fact,
it is easy to see that {\it the trace of any power of the  Lax matrix
$A\lambda+M$ is a Casimir function of the Poisson pencil
$(\Q-\lambda\Pp)$\/}: by definition (\ref{LP}, \ref{Q}),
\BE
(\Q-\lambda\Pp)\ df = [A\lambda+M,\nabla f];
\EE
as already seen, for $f_{\lambda}^{(k)}= 
\frac{1}{k}\tr(A\lambda+M)^k$ one has $\nabla
f_{\lambda}^{(k)}
=(A\lambda+M)^{k-1}$, which obviously commutes with $A\lambda+M$. The same
happens for the Laurent
series expansion of the trace of any half--integer power of
$A+M\lambda^{-1}$. On account of
(\ref{HSP}), {\it all\/} the vector fields generated by the coefficients of
these
Casimir functions (which all mutually commute, by the property mentioned
above) correspond to Lax
equations with spectral parameter:
\BE
X_k = (\Q-\lambda\Pp)\ df^{(k)}_{\lambda} = [A\lambda+M,\nabla
f_{\lambda}^{(k)}].\label{LH}
\EE
In particular, the Manakov equation (\ref{EPM}) corresponds to the first
vector field of the hierarchy
associated to the Casimir function
$f_{\lambda}=\frac{2}{3}\tr(A+M\lambda^{-1})^{3/2}$, for which
$\nabla f_{\lambda}^{(1)} = A^{1/2}\lambda+\Omega$. No comparably simple
and natural connection exists between the other Poisson pencil
$(\R-\lambda\Q)$ and the Lax--Manakov form of the equations.
This setting can be generalized to cover the
cases of more general Lax matrices with spectral parameter on $\gl(r)$, of
the form
$L(\lambda)=A\lambda^k+M_1\lambda^{k-1}+\cdots+M_k$. The general framework
is described in \cite{MagnMagri}, and
will be partially recalled in section (4) below.

So far, we have simply reviewed some results already present in the
literature. Now, some questions
arise naturally. We have described a dynamical system which is
bihamiltonian with respect to two independent PQ structures, $(\Pp,\Q)$ and
$(\Q,\R)$;
is that a pure accident, or it is a common situation?

One can check by explicit computation that the Poisson tensors $\Pp$ and
$\R$ are not only
separately compatible with $\Q$, but also compatible with each other (that
is not obvious,
as compatibility is not a transitive relation). Does it make any sense to
introduce the notion of a
trihamiltonian structure $(\Pp,\Q,\R)$? Would it carry any additional
information not
already contained in either one of the PQ structures, each of which already
allows to characterize
completely the dynamical system and its symmetries?

The vector fields (\ref{LH}) on $\gl(r)$ are indeed trihamiltonian. The
full set of
hamiltonians and vector fields generated by the traces of integers powers
of $A\lambda+M$, fit into
a single ``planar'' diagram (as was first pointed out by M.~Ugaglia
\cite{Ug}), which could be
regarded as the ``trihamiltonian version'' of the Lenard--Magri ``linear''
diagrams (\ref{LMM}):
\BE\label{cry}
\begin{picture}(250,200)
\Scatola{30}{33}{$f_1^3$}
\Scatola{90}{33}{$f_2^3$}
\Scatola{150}{33}{$f_3^3$}
\put(45,48){\NucDir{$\Lie{M}{A^2}$}}
\put(105,48){\NucDir{$\Lie{M^2}{A}$}}
\Scatola{60}{91}{$f_1^2$}
\Scatola{120}{91}{$f_2^2$}
\put(75,106){\NucDir{$\Lie{M}{A}$}}
\Scatola{90}{149}{$f_1^1$}
\multiput(105,164)(30,-58){3}{%
\begin{picture}(30,25)
\put(0,0){\usebox{\Pdir}}
\Scatola{15}{10}{$0$}
\end{picture}
}
\multiput(0,48)(30,58){3}{%
\begin{picture}(30,25)
\put(15,0){\usebox{\Qdir}}
\Scatola{0}{10}{$0$}
\end{picture}
}
\multiput(30,0)(60,0){3}{%
\begin{picture}(15,33)
\put(0,15){\usebox{\Rdir}}
\Scatola{0}{0}{$\cdots$}
\end{picture}
}
\end{picture}
\EE

We have seen above that, for a PQ structure, any linear recursion starting
from a Casimir function of
$Q$ and ending with a Casimir function of $P$ corresponds to the existence of a
$\lambda$--polynomial Casimir function of the Poisson pencil $(P-\lambda
Q)$. Can one find a
``generating polynomial'' for the full trihamiltonian recursion? The answer
is yes: as we shall see
in detail in section (2), if one considers two compatible Poisson pencils
$(\Q-\lambda\Pp)$ and
$(\R-\mu\Pp)$,  one can define a {\it common Casimir function of the two
pencils\/} to be a
bivariate polynomial
$f_{\lambda\mu}=\sum h^i_j\ \lambda^j\mu^i$ such that
\begin{eqnarray}\label{caslm}
(\Q-\lambda\Pp)\ df_{\lambda\mu} &=& 0\nonumber\\
(\R-\mu\Pp)\ df_{\lambda\mu} &=& 0
\end{eqnarray}
for any value of $(\lambda,\mu)$: then its coefficients $h^i_j$ fulfill the
recursion relations
represented in the diagram (\ref{cry}). Later on we will explain why the
construction
of two Poisson pencils, each one with its own spectral parameter, is here
more fruitful than
introducing a two--parameter pencil like $(\Pp-\lambda\Q-\mu\R)$.

Up to this point, the reader might still regard the idea of trihamiltonian
structures as an
artifact of purely academic interest, a mere ``variation on the theme'' of
bihamiltonian
structures. Two results, presented in this article, suggest that the subject is
worth investigating further.

First, the trihamiltonian structure associate to equation (\ref{EPM}) can
be generalized, in quite a nontrivial way, to Lax equations for matrices of
the form
$L(\lambda)=A\lambda^n+M_1\lambda^{n-1}+\cdots+M_n$, which
include several
interesting systems such as the Lagrange top \cite{Ratiu} and the
finite--dimensional
Dubrovin--Novikov reductions of the Gel'fand-Dickey soliton hierarchies
\cite{DubNov}. Indeed, the
generalization of the pencil $(\Q-\lambda\Pp)$ to the direct sum of
$n$ copies of $\gl(r)$ was
already described in \cite{MagnMagri}, but to our knowledge it is an
entirely new result that also
the Morosi--Pizzocchero bracket is a particular case of a more general
structure existing on
$\gl(r)^n$, a Poisson tensor $\R$ which turns out to be {\it
quadratically\/}
dependent on the dynamical variables $M_i$, apart for the linear case
$n=1$ already discussed.

The second striking fact is that for these trihamiltonian structures on
$\gl(r)^n$ there always exists a common Casimir function of the two
pencils, which (for a generic
choice of the matrix $A$) has the property that its coefficients form a
{\it maximal\/} set of {\it
independent} hamiltonians in involution (we stress that, in contrast, the
recursion diagram for the traces of the powers
of the Lax matrix includes infinitely many hamiltonians, and one has to
single out
a finite subset of independent first integrals). This miraculous Casimir
function
is nothing but the characteristic determinant of the Lax matrix,
\BE
f_{\lambda\mu} = \det | L(\lambda)- \mu\identity |. \label{cs}
\EE
The corresponding recursion diagram features a sort of ``fundamental
molecule'', a ``fingerprint'' associated to the
trihamiltonian structure $(P,Q_1,Q_2)$. For instance, the following diagram
corresponds to the trihamiltonian recursion on the
algebra $\gl(3)$:
\BE
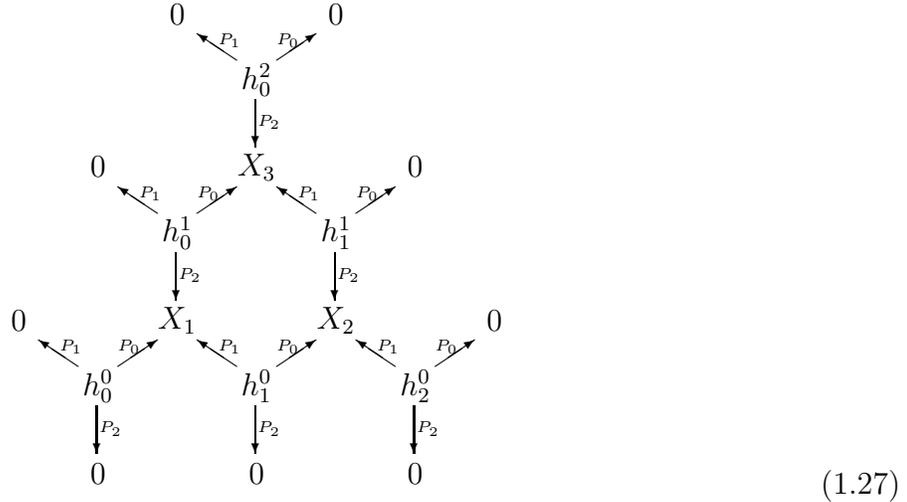
\label{cry1}
\begin{picture}(250,180)
\Scatola{30}{33}{$h^0_0$}
\Scatola{90}{33}{$h^0_1$}
\Scatola{150}{33}{$h^0_2$}
\put(45,48){\NucDir{$X_1$}}
\put(105,48){\NucDir{$X_2$}}
\Scatola{60}{91}{$h^1_0$}
\Scatola{120}{91}{$h^1_1$}
\put(75,106){\NucDir{$X_3$}}
\Scatola{90}{149}{$h^2_0$}
\multiput(30,0)(60,0){3}{%
\begin{picture}(15,33)
\put(0,15){\usebox{\Rdir}}
\Scatola{0}{0}{$0$}
\end{picture}
}
\multiput(0,48)(30,58){3}{%
\begin{picture}(30,25)
\put(15,0){\usebox{\Qdir}}
\Scatola{0}{10}{$0$}
\end{picture}
}
\multiput(165,48)(-30,58){3}{%
\begin{picture}(30,25)
\put(0,0){\usebox{\Pdir}}
\Scatola{15}{10}{$0$}
\end{picture}
}
\end{picture}
\EE
while (\ref{cry2}) is the ``molecule'' of the trihamiltonian structure on
$\gl(2)^2$:
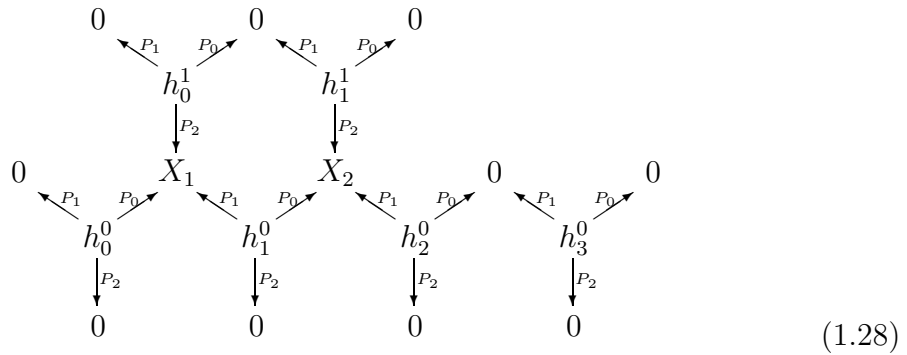
\BE\label{cry2}
\begin{picture}(250,140)
\Scatola{30}{33}{$h^0_0$}
\Scatola{90}{33}{$h^0_1$}
\Scatola{150}{33}{$h^0_2$}
\Scatola{210}{33}{$h^0_3$}
\put(45,48){\NucDir{$X_1$}}
\put(105,48){\NucDir{$X_2$}}
\Scatola{60}{91}{$h^1_0$}
\Scatola{120}{91}{$h^1_1$}
\multiput(165,48)(-90,58){2}{%
\begin{picture}(45,25)
\put(0,0){\usebox{\Pdir}}
\Scatola{15}{10}{$0$}
\put(30,0){\usebox{\Qdir}}
\end{picture}
}
\multiput(30,0)(60,0){4}{%
\begin{picture}(15,33)
\put(0,15){\usebox{\Rdir}}
\Scatola{0}{0}{$0$}
\end{picture}
}
\multiput(0,48)(30,58){2}{%
\begin{picture}(30,25)
\put(15,0){\usebox{\Qdir}}
\Scatola{0}{10}{$0$}
\end{picture}
}
\multiput(225,48)(-90,58){2}{%
\begin{picture}(30,25)
\put(0,0){\usebox{\Pdir}}
\Scatola{15}{10}{$0$}
\end{picture}
}
\end{picture}
\EE
The general form of the ``fundamental molecule'' for $\gl(r)^n$ is
given in
section (4) as
fig.1.

Although it is a well known fact that the coefficients of the characteristic
polynomial are in involution with respect to the usual Lie--Poisson
bracket, in the
bihamiltonian framework there was no apparent reason to introduce a bivariate
polynomial $f_{\lambda\mu}$ in connection with the Lenard--Magri recursion.
For a
trihamiltonian structure, instead, it is quite natural to consider this
object,
and the characteristic polynomial of a Lax matrix becomes just a particular
case of it,
in exactly the same way as Lax equations with spectral parameter are a
particular case of Hamilton equations, for the appropriate Poisson pencil
(\ref{HSP}).

This opens a very interesting perspective. The characteristic equation
\BE
\det | L(\lambda)- \mu\identity | = 0, \label{spec}
\EE
regarded as a polynomial equation for $(\lambda,
\mu)\in\Complessi^2$ defines the well--known {\it spectral curve}, i.e.~the
starting point for the
algebro--geometric methods of linearisation \cite{Grif} \cite{Van}. In the
trihamiltonian framework, as we have
seen, the characteristic determinant naturally occurs as the fundamental
Casimir function of two pencils: yet this does not explain why the roots
$(\lambda_i,\mu_i)$ of eq.~(\ref{spec}) should play any role at all. Now
comes a third
surprise: a fairly general construction presented in section (3) shows that
the equation
$f_{\lambda\mu}=0$ is the keystone for the construction of canonical
separation coordinates for
trihamiltonian systems.

This result essentially derives from an observation by E. Sklyanin
\cite{Skly}. On algebro--geometric
grounds, Sklyanin has found a ``magic recipe''({\it
``Take the poles of the properly normalized Baker-Akhiezer function and the
corresponding
eigenvalues of the Lax operator''}), which essentially amounts to finding
the common roots of
(\ref{spec}) and of suitable minors (or linear combination of minors) of
the characteristic matrix
$L(\lambda)-\mu\identity$. In the examples considered by Sklyanin, the new
variables $(\lambda_i,
\mu_i)$ defined in this way turn out to be canonical with respect to a
suitable Poisson bracket; by
direct consequence of eq.~(\ref{spec}), all the hamiltonians defined as the
(nonconstant) coefficients of $f_{\lambda\mu}$ are then separable in the
coordinates $(\lambda_i,
\mu_i)$. However, Sklyanin himself remarks that {\it ``generally speaking,
there
is no guarantee that one
obtains the canonical Poisson brackets [..] The key words in the above
recipe are `the properly
normalized'. The choice of the proper normalization can be quite
nontrivial, and for some integrable
models the problem remains unsolved''.\/}

Independently of Sklyanin's approach, Magri and his collaborators
\cite{Magri2}\cite{Magri2b} have recently shown
that  given (i) a PQ structure, (ii) a complete family of commuting
hamiltonians defined by the
Casimir functions of the Poisson pencil, and (iii) a set of vector fields,
suitably normalized on the
hamiltonians previously introduced, which preserve the Poisson tensor $P$
but do not belong to its
image (geometrically speaking, they should be transversal to the
symplectic leaves of $P$),
then one can define by projection (under some additional conditions on the
vector fields) a reduced,
kernel--free bihamiltonian structure; for this new PQ structure, a set of {\it
Darboux--Nijenhuis canonical coordinates\/} can be obtained by a
constructive procedure,
and the original hamiltonians (properly reduced) turn out to be all
simultaneously separable in these
coordinates.

The theoretical interest of both constructions is largely beyond the
concrete applicability of these procedures. As a matter of fact, while
Sklyanin's
recipe lacks a general, theoretically--grounded rule to find the key
element (the
normalization of the BA function, or equivalently the proper linear combination
of minors of the Lax matrix which should vanish), in Magri's theory there is no
practical recipe to construct
systematically sets of transversal vector fields fulfilling the necessary
requirements. In both
approaches, moreover, the final construction of separation coordinates
involves finding the roots
of polynomial equations, which even for rather simple
examples turn out to be of order higher than three.

As we show in this article, Magri's procedure can be adapted to the
trihamiltonian setup,
without loosing its geometric elegance, and actually making the theory even
simpler and more symmetric (although less general). In this framework, the
central role of the
``generalized spectral equation''
$f_{\lambda\mu}=0$ becomes clear. Moreover, for the trihamiltonian
structures that we are
introducing on the spaces $\gl(r)^n$, we have found a systematic way
to produce explicitly the
required transversal vector fields, and we will show that the associated
Darboux--Nijenhuis coordinates
are exactly the roots of suitable combinations of minors of the 
characteristic matrix,
much alike Sklyanin's coordinates; in this way, we provide for this 
class of systems the missing
element in both Sklyanin's and Magri's prescriptions for
the construction of separation variables. In addition, we show that our
framework makes available a
different strategy, which yields the {\it inverse transformation\/}
(i.e.~the matrix elements of the
Lax operator as functions of the separation variables) by solving only a
system of
\emph{linear} algebraic equations, thus bypassing the problem of finding
roots of
higher--order polynomials.

Let us quote another important remark by Sklyanin
\cite{Skly}: {\it ``Separation of
variables, understood generally enough, could be the most universal tool to
solve integrable models
[...] the standard construction of the action--angle variables from the
poles of the Baker--Akhiezer
function can be interpreted as a variant of separation of variables, and
moreover, for many
particular models it has a direct quantum counterpart''.\/} Therefore, a
satisfactory hamiltonian
setup for Sklyanin's construction is likely to provide a link between
hamiltonian and
algebro--geometric integrability. In this sense, the equation
$f_{\lambda\mu}=0$ should deserve some additional interest, as it points
towards a
generalisation of the notion of spectral curve not relying on Lax
representations.

The article is organized as follows: in section (2) we recall, as
synthetically as possible, some
facts about bihamiltonian structures which are necessary for the subsequent
discussion; then, we
present a theoretical setting of our class of trihamiltonian structures. In
section (3) we discuss the
general method of construction of separation variables, i.e.~the
trihamiltonian version of
Magri's construction. We present in detail the proofs of some relevant
propositions providing the
theoretical background for all applications of our framework; furthermore,
we show how the components of
all relevant objects (Poisson structures, common Casimir function,
transversal vector fields, etc.)
look like in Darboux--Nijenhuis coordinates; this will be used in section
(4) to reconstruct the
coordinate transformation. The fourth and last
section is devoted to the application to Lax equations with spectral
parameter on
$\gl(r)^n$; here we
simply list the ``ingredients of the recipe'' without a general 
proof; this section is
intended to present a concrete outcome in just enough
detail to motivate the reader to deal with the theoretical 
construction of section (3). 

Throughout the article no attempt is made to present
a geometric characterisation, or classification, of the trihamiltonian structures possessing
the specific features considered. In particular, we deliberately avoid to encompass these features into a single
definition of ``special trihamiltonian structure''. In our discussion,
the basic structure involved is sometimes presented as a {\it triple} of compatible Poisson tensors, but more often as a
{\it pair} of Possion pencils; the assumptions on this basic structure vary according to the context. In sect.(2) we
just reconsider the notion of ``trihamiltonian recursion'' associated with a common Casimir function of two Poisson
pencils, without imposing particular conditions on the Poisson tensors besides their mutual compatibility; in such
generality, indeed, nothing ensures that a common Casimir function exist at all (we will present a counterexample). In
sect.(3.1) we show how to define a pair of Nijenhuis tensors using a set of vectorfields, which should fulfill a
number of requirements: these Nijenhuis tensors are related with a particular system of Darboux--Nijenhuis coordinates,
and do {\it not} act as recursion operators for the trihamiltonian iteration described previously (throughout this
subsection, the existence of a common Casimir function is irrelevant). In section (3.2) the two objects, i.e.~the common
Casimir function (that we now require to be {\it complete} in a suitable sense) and the set of transversal vectorfields,
are eventually combined together to construct a set of bivariate polynomials $S_\alpha(\lambda,\mu)$: then, a number of
additional hypotheses are introduced to obtain the main result, i.e.~that the common roots of these polynomials are
Darboux--Nijenhuis coordinates, in which the Hamiltonians occurring as coefficients in the common Casimir function are
separated (in the sense of Sklyanin). Hence, the set of conditions to be imposed on the basic structure depends on
whether one looks for the mere existence of iterated trihamiltonian vectorfields, or for a possible explicit construction
of separation coordinates. In conclusion, we feel that at the present stage of understanding of the matter presented
herein, one could hardly find a simple and general definition which might be regarded as truly fundamental.  The aim of
the article is rather to display a number of nontrivial and rigorous (if not yet complete) arguments in favour of
further investigations in this direction.

\setcounter{equation}{0}
\section{From bi-- to trihamiltonian structures}
\subsection{Poisson pencils}
As was already done (\ref{pbra}) in the introductory section, we shall
represent a Poisson
bracket \mbox{$\Pois{\cdot}{\cdot}$} on a manifold $\cal M$ by means of a
contravariant
antisymmetric tensorfield $P$, according to
\BE\label{poisbra}
\langle dg, Pdf\rangle = \Pois{f}{g} .
\EE
The names \emph{Poisson structure} or \emph{hamiltonian structure} are
equivalently used, as is
commonly done, to denote both the tensor $P$ and the algebra of
differentiable functions on $\cal M$
with the bilinear operation defined by the corresponding Poisson bracket.
Of course, a
contravariant antisymmetric tensorfield $P$ defines a hamiltonian
structure only if the bracket
(\ref{poisbra}) obeys the Jacobi identity; this condition corresponds to a
differential
identity on the components of $P$.

In most of our applications, the tensor
$P$ will not be of maximal rank; thus, the subalgebra of functions which are in
involution with any other function may include non-constant functions, the
\emph{Casimir functions}. The Casimir functions are constant of motion for any
\emph{hamiltonian vectorfield},
i.e.~for any vectorfield being the image of a closed one--form through
the Poisson tensor $P$. Therefore, any trajectory of any possible hamiltonian
system on that phase space lies entirely on a common level set of all the
Casimir functions. Generically, such
a level set is a submanifold, the dimension of which equals the rank of the
Poisson tensor. Upon reduction to
any of these submanifolds, the Poisson tensor becomes invertible and
therefore defines a \emph{symplectic
structure}. For this reason, the common level sets (for regular values) of
the Casimir functions are called
\emph{symplectic leaves}.
In contrast with the case of symplectic manifolds, the Lie derivative the
Poisson tensor can vanish along the flow of a given vectorfield $X$,
\BE
{\cal L}_X(P) = 0 , \label{symm}
\EE
without $X$ being even locally hamiltonian: the condition (\ref{symm}) can be
fulfilled also by
vectorfields which do not belong to the image of $P$, and are therefore
transversal to the
symplectic leaves. Such vectorfields will be called \emph{weakly
hamiltonian}. They play an
important role in the sequel.

Bihamiltonian structures were first introduced by Magri in \cite{Magri1}.
\begin{Def}
Two Poisson tensors $P$ and $Q$ on a manifold $\cal M$ are told to
be \emph{compatible} if any linear combination of the two tensors is again a
Poisson tensor.
\end{Def}
In such situation one can find vectorfields which are hamiltonian with
respect to both structures, i.e.~\emph{bihamiltonian vectorfields}. In this
article, we borrow from
the bihamiltonian theory the following facts:

(i) Two hamiltonians are associated to a single bihamiltonian vectors field
$X=Pdh=Qdk$. Then, one can define two other vectorfields, namely $Qdh$
and $Pdk$.
In some cases, these turn out to be bihamiltonian as well, and the
procedure can be iterated
yielding a \emph{Magri--Lenard hierarchy} of bihamiltonian vectorfields,
as in
(\ref{lenard}).

(ii) Once a Magri--Lenard hierarchy has been constructed, all the
vectorfields belonging to it are
mutually commuting, and all their hamiltonians are in involution with
respect to both $P$ and $Q$.

(iii) There are basically two ways to produce such hierarchies: if at least
one of the Poisson
tensors (say,
$P$) is nondegenerate, then one can introduce the \emph{recursion operator}
(or \emph{Nijenhuis
tensor})
\BE
N = Q \cdot P^{-1};\label{nij}
\EE
One can prove \cite{Magri3} that for any bihamiltonian vectors field $X$,
the vectors field $NX$ is also
bihamiltonian, so the hierarchy can be produced by iterated application of
the $(1,1)$ tensors field
$N$. Alternatively (for instance, if both Poisson tensors are degenerate),
one can look for
Casimir functions of the Poisson pencil
$(Q-\lambda P)$, as already described in the Introduction.

The classical proof of the involutivity property, which is the most
relevant to our purposes, is so
simple and elegant that we reproduce it here (further details can be found
in \cite{Magri5}, \cite{Magn}):

\begin{Prop}{}\label{prop}
Let $f_{\lambda}$ and $g_{\lambda}$ be two Casimir functions of the Poisson
pencil $(Q-\lambda P)$.
Assume that $f_{\lambda}$ and $g_{\lambda}$ are expanded in power series in
the parameter $\lambda$
of the pencil, $f_{\lambda}=\sum_{i=0} f_i\lambda^i$ and
$g_{\lambda}=\sum_{i=0} g_i\lambda^i$.
Then $\Pois{f_j}{f_k}=\Pois{g_j}{g_k}=\Pois{f_j}{g_k}=0$ for all $j,k$:
this holds for both brackets
$\Pois{\ }{\ }_{_P}$ and $\Pois{\ }{\ }_{_Q}$ associate to $P$ and to $Q$
respectively.
\begin{Dim}
The conditions $(Q-\lambda P)\ df_{\lambda}=0$ and $(Q-\lambda P)\
dg_{\lambda}=0$ are equivalent to
$P df_i=Q df_{i+1}$ and $P dg_i=Q dg_{i+1}$. Moreover, one should have $Q
df_0=0$ and $Q dg_0=0$,
i.e.~the lowest--order coefficients of both expansions should be Casimir
function for $Q$.
One can assume $j<k$,
without loss of generality. From the definition (\ref{poisbra})
$\Pois{f_j}{f_k}_{_P}=\Pois{f_{j+1}}{f_k}_{_Q}=\Pois{f_{j+1}}{f_{k-1}}_{_P}$.
Whenever $k-j$ is even,
applying repeatedly the equality one finds that
$\Pois{f_j}{f_k}_{_P}=\Pois{f_{r}}{f_{r}}_{_P}=0$,
with $r=(k-j)/2$; otherwise, after $(k-j)$ steps one finds
$\Pois{f_j}{f_k}_{_P}=\Pois{f_{k}}{f_{j}}_{_P}$, which proves the statement
by the antisymmetry of
the Poisson bracket. The same holds for $\Pois{g_j}{g_k}_{_P}$, and for the
other bracket $\Pois{\
}{\ }_{_Q}$. Furthermore, applying the same iterative argument one finds
$\Pois{f_j}{g_k}_{_Q}=\Pois{f_{j+k}}{g_{0}}_{_Q}$, and the latter bracket
vanishes because
$g_0$ is a Casimir function for $Q$. This proves that
$\Pois{f_j}{g_k}_{_Q}=0$ for all $j,k$. Since
$\Pois{f_j}{g_k}_{_P}=\Pois{f_{j+1}}{g_k}_{_Q}$, one has
$\Pois{f_j}{g_k}_{_P}=0$ as well.
\end{Dim}
\end{Prop}

\subsection{Casimir functions and trihamiltonian vector fields}

Assume that a manifold $\mathcal{M}$ is endowed with three Poisson
tensors $\Pp$, $\Q$ and $\R$, pairwise compatible. A natural question is
whether a set of vectorfields that are hamiltonian with respect to all three
structures can be generated by the coefficients of some ``generating
function'',
analogous to the Casimir function $f_{\lambda}$ above, and whether the
corresponding hamiltonians
would then be automatically in involution.

One might believe that the obvious generalization of the setting
just described would
consist in introducing a two--parameter Poisson pencil
\BD
\Pp - \lambda \Q - \mu \R
\ED
and seeking for its Casimir functions. Unfortunately, the coefficients of
the Taylor series in the two
parameters
$(\lambda,\mu)$
do not fit into any useful recursion relation: from the
Casimir equation
\BD
(\Pp -
\lambda\Q - \mu\R)\ df_{\lambda\mu} = 0  \qquad
\mathrm{for} \qquad
f_{\lambda,\mu} = \sum_{i,j=0}^{\infty} h_i^j \lambda^i \mu^j
\ED
one gets the relations
\begin{eqnarray*}
\Pp\, dh_0^0 & = & 0 \\ \Pp \, dh_{i+1}^0 & = & \Q \, dh_i^0 \hskip80pt
i = 0, 1, \ldots \\
\Pp \, dh_0^{j+1} & = & \R \, dh_0^j \hskip80pt j = 0, 1, \ldots \\
\Pp \, dh_{i+1}^{j+1} & = & \Q \, dh_i^{j+1} + \R \, dh_{i+1}^j
\hskip20pt i,j = 0, 1, \ldots
\end{eqnarray*}
which neither provide trihamiltonian vector fields nor force
the functions $f_i^j$ to be in involution.

Let us consider instead a function $f_{\lambda\mu}$
that is simultaneously a Casimir function of the two distinct pencils
\mbox{$(\Q - \lambda\Pp)$} and
\mbox{$(\R - \mu\Pp)$}:
\BD
(\Q - \lambda\Pp) \, df_{\lambda\mu} = 0 \,,\qquad (\R - \mu\Pp) \,
df_{\lambda\mu} = 0 \,.
\ED
In this case we actually obtain the following relations:
\begin{eqnarray*}
\Q \, dh_0^j & = & 0 \hskip70pt j = 0, 1, \ldots \\
\Pp \, dh_i^j & = &
\Q \, dh_{i+1}^j \hskip35pt i,j = 0, 1, \ldots \\
\R \, dh_i^0 & = & 0 \hskip70pt i = 0, 1, \ldots \\
\Pp \, dh_i^j & = &
\R \, dh_i^{j+1} \hskip35pt i,j = 0, 1, \ldots
\end{eqnarray*}
graphically:
\BE\label{infcry}
\begin{picture}(300,230)
\Scatola{30}{33}{$h_0^0$}
\Scatola{90}{33}{$h_1^0$}
\Scatola{150}{33}{$h_2^0$}
\Scatola{210}{33}{$\cdots$}
\put(45,48){\NucDir{$X_1$}}
\put(105,48){\NucDir{$X_2$}}
\put(165,48){\NucDir{$X_3$}}
\Scatola{60}{91}{$h_0^1$}
\Scatola{120}{91}{$h_1^1$}
\Scatola{180}{91}{$h_2^1$}
\Scatola{240}{91}{$\cdots$}
\put(75,106){\NucDir{$Y_1$}}
\put(135,106){\NucDir{$Y_2$}}
\put(195,106){\NucDir{$Y_3$}}
\Scatola{90}{149}{$h_0^2$}
\Scatola{150}{149}{$h_1^2$}
\Scatola{210}{149}{$\cdots$}
\put(105,164){\NucDir{$Z_1$}}
\put(165,164){\NucDir{$Z_2$}}
\Scatola{120}{207}{$\cdots$}
\Scatola{180}{207}{$\cdots$}
\multiput(30,0)(60,0){3}{%
\begin{picture}(15,33)
\put(0,15){\usebox{\Rdir}}
\Scatola{0}{0}{$0$}
\end{picture}
}
\multiput(0,48)(30,58){3}{%
\begin{picture}(30,25)
\put(15,0){\usebox{\Qdir}}
\Scatola{0}{10}{$0$}
\end{picture}
}
\end{picture}
\EE
\noindent the vectorfields \mbox{$\Pp \, dh_i^j =  \Q \, dh_{i+1}^j
=  \R \, dh_i^{j+1}$} are clearly trihamiltonian.

Notice that it is possible to find a common
Casimir function which can be (formally) expanded in a Taylor series with
respect to the two parameters $\lambda$ e $\mu$ only if both $\Q$ and $\R$
are degenerated
Poisson tensors: in fact, for a fixed power of $\lambda$, the lowest--order
coefficient in $\mu$
must be a Casimir function of $\R$, while the lowest--order coefficient in
$\lambda$ for any fixed
power of $\mu$ must be a Casimir function of $\Q$. If, moreover, also $\Pp$
is degenerate, then it
is possible to find Casimir functions which are polynomials in $\lambda$
and $\mu$. For such
functions the recursion diagram is finite.

In analogy with the bihamiltonian case, one has:

\begin{Prop}{}
Given a common Casimir function \mbox{$f_{\lambda\mu} = \sum h_i^j \lambda^i
\mu^j$} of two compatible Poisson pencils \mbox{$\Q -\lambda\Pp$} and \mbox{$\R
-\mu\Pp$}, all the coefficients
$h_i^j$ are in mutual involution with respect to all three Poisson
brackets.
\begin{Dim}
For any $i,j$ the functions \mbox{$h_\lambda^{(i)} = \sum h_k^i
\lambda^k$} and
\mbox{$h_\lambda^{(j)} = \sum h_k^j \lambda^k$}, i.e.~the total
coefficients of $\mu^i$ and $\mu^j$
in the expansion of $f_{\lambda\mu}$, are Casimir functions of the Poisson
pencil
\mbox{$(\Q -\lambda\Pp)$}; then, by proposition (\ref{prop}) all their
coefficients $h_i^j$ are in
involution with respect to both $\Pp$ and $\Q$. On the other hand, the
functions
\mbox{$h_\mu^{(i)} = \sum
h_i^k \mu^k$} and \mbox{$h_\mu^{(j)} = \sum h_j^k \mu^k$} are
Casimir functions of the other pencil
\mbox{$\R -\mu \Pp$}, hence  $h_i^j$ are in involution also with
respect to $\R$.
\end{Dim}
\end{Prop}

We remark that there are other possible ways to extend the bihamiltonian
framework to the
case in which there are more than two compatible Poisson structures. The
idea of a trihamiltonian
vector field was already considered, for example, in \cite{Oevel} and
\cite{MorPizz2}), where the three structures $P$, $Q$ and $S$ were however
assumed to produce the
iteration
$
P \D h_i = Q \D h_{i+1} = S \D h_{i-1}
$.
In this approach, the third structure only supplies an additional relation
which links vectorfields already belonging to the same Magri--Lenard
hierarchy;
in our framework, the third
structure acts instead as a bridge
linking different bihamiltonian hierarchies, and so allows to collect a
greater number of function
in involution in a single objet: the common Casimir function.

In our setup, the structure $\Pp$ seem to play a distinguished role
with respect to $\Q$ and $\R$. As a
matter of fact, it is
easy to figure out how to include in the picture also the Poisson pencil
built from $\Q$ and
$\R$, but the $\Q$--$\R$ recursion is \emph{already} included in the
diagram (\ref{infcry}), and
introducing a third pencil would be redundant. In the recursion diagram,
all structures appear on equal
footing; on the other hand, in the applications that we have in mind there
\emph{is} a distinguished structure, so the ``symmetry breaking'' caused by
the choice of two
pencils is significant.

We stress that, given three Poisson structures $\Pp$, $\Q$ and $\R$, mutually compatible and such that
the two pencils \mbox{$\Q-\lambda\Pp$}, \mbox{$\R-\mu\Pp$} both admit 
Casimir functions, a common Casimir function as required in Prop.(2.2) may not exist at all. An obvious necessary condition is that
at each point $x$ of the phase space $\mathcal{M}$, and for generic values of the spectral parameters $(\lambda,\mu)$, the two
subspaces $ker(\Q-\lambda\Pp)$ and $ker(\R-\mu\Pp)$ should have a nontrivial intersection in $T^*_x\mathcal{M}$.
For instance, let us consider the space $\Reali^5$, with coordinates
$\{x_1,x_2,x_3,x_4,x_5\}$, endowed with the three Poisson tensors:
\BD
\begin{array}{c}
\scriptsize \Pp = \left(
\begin{array}{ccccc}
0 & 0 & 1 & 0 & 0 \\
0 & 0 & 0 & 1 & 0 \\
-1 & 0 & 0 & 0 & 0 \\
0 & -1 & 0 & 0 & 0 \\
0 & 0 & 0 & 0 & 0
\end{array}
\right) \\ \noalign{\vskip10pt}
\scriptsize \Q = \left(
\begin{array}{ccccc}
0 & 0 & 0 & 1 & 0 \\
0 & 0 & 0 & 0 & 1 \\
0 & 0 & 0 & 0 & 0 \\
-1 & 0 & 0 & 0 & 0 \\
0 & -1 & 0 & 0 & 0
\end{array}
\right) 
\qquad
\R = \left(
\begin{array}{ccccc}
0 & 0 & 0 & 1 & 0 \\
0 & 0 & 0 & 0 & 0 \\
0 & 0 & 0 & 0 & 1 \\
-1 & 0 & 0 & 0 & 0 \\
0 & 0 & -1 & 0 & 0
\end{array}
\right)
\end{array}
\ED
Although the pencils \mbox{$\Q-\lambda\Pp$} and \mbox{$\R-\lambda\Pp$} admit the Casimir functions \mbox{$x_3+\lambda
x_4+\lambda^2 x_5$} and
\mbox{$x_2+\mu x_1-\mu^2 x_5$}, respectively, it is easy to check that a common Casimir function for both pencils does not exist.

The simple local geometry of our trihamiltonian structures may be clarified
by an example. Let us consider
the ``fundamental molecule'' (\ref{cry1}). The lowest
dimension in which this diagram can be realised is 9. In
fact, the diagram includes
six functions, that we assume to be independent. Since the three
vectorfields in the diagram
commute, by Frobenius' theorem there exists a coordinate system in which
they coincide with
coordinate vectorfields: $X_i\equiv\frac{\partial}{\partial x^i}$ for
$i=1,2,3$. The diagram shows that the hamiltonian $h^0_0$ is
$\Pp$--conjugate to $x^1$, while
$h^0_1$ and
$h^1_0$ are
$\Pp$--conjugate to
$x^2$ and $x^3$ respectively. The other hamiltonians $h^0_2$, $h^1_1$ and
$h^2_0$ are Casimir
functions for $\Pp$. Therefore, the 9 functions $x^i$ and $h^i_j$ should be
functionally
independent, and locally form a coordinate system: let $x^4\equiv h^0_0$,
$x^5\equiv h^0_1$,
$x^6\equiv h^1_0$,
$x^7\equiv h^0_2$,
$x^8\equiv h^1_1$ and $x^9\equiv h^2_0$. It can be read directly from the
diagram (\ref{cry1}) that
in these coordinates the three tensors $\Pp$, $\Q$ and $\R$ have the
following matrix
components:
\BD
\matrix{
\tiny \Pp = \pmatrix{
0 & 0 & 0 & 1 & 0 & 0 & 0 & 0 & 0 \cr
0 & 0 & 0 & 0 & 1 & 0 & 0 & 0 & 0 \cr
0 & 0 & 0 & 0 & 0 & 1 & 0 & 0 & 0 \cr
-1 & 0 & 0 & 0 & 0 & 0 & 0 & 0 & 0 \cr
0 & -1 & 0 & 0 & 0 & 0 & 0 & 0 & 0 \cr
0 & 0 & -1 & 0 & 0 & 0 & 0 & 0 & 0 \cr
0 & 0 & 0 & 0 & 0 & 0 & 0 & 0 & 0 \cr
0 & 0 & 0 & 0 & 0 & 0 & 0 & 0 & 0 \cr
0 & 0 & 0 & 0 & 0 & 0 & 0 & 0 & 0 }
  \cr\noalign{\vskip10pt}
\tiny \Q = \pmatrix{
0 & 0 & 0 & 0 & 1 & 0 & 0 & 0 & 0 \cr
0 & 0 & 0 & 0 & 0 & 1 & 0 & 0 & 0 \cr
0 & 0 & 0 & 0 & 0 & 0 & 1 & 0 & 0 \cr
0 & 0 & 0 & 0 & 0 & 0 & 0 & 0 & 0 \cr
-1 & 0 & 0 & 0 & 0 & 0 & 0 & 0 & 0 \cr
0 & -1 & 0 & 0 & 0 & 0 & 0 & 0 & 0 \cr
0 & 0 & -1 & 0 & 0 & 0 & 0 & 0 & 0 \cr
0 & 0 & 0 & 0 & 0 & 0 & 0 & 0 & 0 \cr
0 & 0 & 0 & 0 & 0 & 0 & 0 & 0 & 0 }
  \qquad
\R = \pmatrix{
0 & 0 & 0 & 0 & 0 & 1 & 0 & 0 & 0 \cr
0 & 0 & 0 & 0 & 0 & 0 & 0 & 1 & 0 \cr
0 & 0 & 0 & 0 & 0 & 0 & 0 & 0 & 1 \cr
0 & 0 & 0 & 0 & 0 & 0 & 0 & 0 & 0 \cr
0 & 0 & 0 & 0 & 0 & 0 & 0 & 0 & 0 \cr
-1 & 0 & 0 & 0 & 0 & 0 & 0 & 0 & 0 \cr
0 & 0 & 0 & 0 & 0 & 0 & 0 & 0 & 0 \cr
0 & -1 & 0 & 0 & 0 & 0 & 0 & 0 & 0 \cr
0 & 0 & -1 & 0 & 0 & 0 & 0 & 0 & 0 }
}
\ED
The fact that three \emph{independent} Poisson tensor can be simultaneously
put in canonical form is
possible only because they are all degenerate (in the realization of minimal
dimension, they ought to have the same rank), and their
symplectic foliations are
different. Having anticipated that the diagram (\ref{cry1}) corresponds to
the trihamiltonian
structure of $\gl(3)$, we have in fact shown that the latter trihamiltonian
space admits local
\emph{multicanonical} coordinates. On the other hand, although six
of the multicanonical
coordinates coincide with the coefficients of the characteristic
determinant of the Lax matrix
$A\lambda+M$, the first three coordinates can be found only upon explicit
integration of the
dynamical system. Analogous considerations hold for the ``fundamental
molecule'' of any space
$\gl(n)^\kappa$, and actually for any finite trihamiltonian recursion
diagram (under the
assumption that all the hamiltonians are independent, which is generically
true for
$\gl(n)^\kappa$). The lowest dimension to accommodate a trihamiltonian
structure
admitting multicanonical coordinates is 4 (the ``fundamental molecule''
contains just one
vectors field and three hamiltonians); an example is the algebra $\gl(2)$. The
reader can easily find out the multicanonical form of the three Poisson
tensors generating the
$\gl(2)^2$ diagram (\ref{cry2}).

Another type of coordinates, the separation coordinates, can instead be
obtained explicitly in an
alternative way, which does not require the integration of the
vectorfields. This is explained in
the next section.

\setcounter{equation}{0}
\section{Separation of variables}

\subsection
{Darboux-Nijenhuis coordinates}

In this section we shall adapt to the trihamiltonian framework the
construction of separation variables
proposed by Falqui, Magri and Pedroni in \cite{Magri2}\cite{Magri2b}.

The basic notion involved in their construction is the definition of
\emph{Darboux-Nijenhuis coordinates} \cite{Magri3} \cite{Marsico}. Consider a
bihamiltonian structure $PQ$ on a
manifold
$\mathcal{M}$, with $\dim(\mathcal{M})=2m$, such that at least one of the
Poisson tensors (say, $P$) is
nondegenerate. Let $N$ be the recursion operator defined by 
(\ref{nij}); following Magri
\cite{Magri3}, we say that $\mathcal{M}$ is endowed with a {\it 
Poisson-Nijenhuis structure.}
We shall assume that $P$ and $Q$ are such that at generic point of
$\mathcal{M}$, the recursion tensor
$N$ has $m$ distinct (double) eigenvalues $\lambda_i$.
If moreover the $m$ eigenvalues $\lambda_i$ are functionally 
independent when regarded
as functions on $\mathcal{M}$, then it has been proved \cite{Marsico} that (at
least locally) other $m$ functions $\mu_i$ exist such that:

(i) the functions $(\lambda_i,\mu_i)$ form a system of coordinates on
$\mathcal{M}$;

(ii) in this coordinate system, the Poisson tensor $P$ is in canonical
form, i.e.
$\Pois{\lambda_i}{\mu_j}_{_P}=\delta_{ij}$ and
$\Pois{\lambda_i}{\lambda_j}_{_P}=\Pois{\mu_i}{\mu_j}_{_P}=0$, and the
recursion tensor $N$ is
diagonal.

The coordinates $(\lambda_i,\mu_i)$ are called
\emph{Darboux-Nijenhuis coordinates}. The property (ii) completely
determines the
$Q$--Poisson brackets of the coordinates:
\mbox{$\Pois{\lambda_i}{\mu_i}_{_Q}=\lambda_i$} and
$\Pois{\lambda_i}{\mu_j}_{_Q}=\Pois{\lambda_i}{\lambda_j}_{_Q}=\Pois{\mu_i}{\mu_
j}_{_Q}=0$
for $i\not= j$.

Suppose $h_i$ to be a set of $m$ hamiltonians, independent and in
involution with respect to $P$ and $Q$ (although not necessarily generated
by Magri--Lenard recursion). Falqui, Magri and Pedroni
\cite{Magri2}\cite{Magri2b} have recently found an intrinsic coupling condition with the
recursion operator, ensuring that all the
functions
$h_i$ are separable in the Darboux-Nijenhuis coordinates.

In the case of a degenerate $PQ$ structure, one can sometimes perform a
reduction onto a symplectic leaf of $P$, by projection along appropriate
transversal vectorfields. This allows one
to compute explicitly not only the
coordinates $\lambda_i$ (which are the roots of the characteristic
polynomial of $N$, or rather of its
\emph{minimal polynomial}), but also the other coordinates $\mu_i$, as the
values taken by a suitable
polynomial $p(\lambda)$ after the substitutions $\lambda=\lambda_i$ (for
the construction of $p(\lambda)$,
the exact statements and the proofs we refer the reader to \cite{Tondo}).

For a trihamiltonian structure
$(\Pp,\Q,\R)$ with a nondegenerate Poisson
tensor
$\Pp$, one is naturally led to introduce \emph{two} recursion operators,
\BE
\Nq = \Q\cdot \Pp^{-1}\qquad\mathrm{and}\qquad\Nr = \R\cdot
\Pp^{-1},\label{nij2}
\EE
so obtaining a compatible pair of Poisson-Nijenhuis structures. In 
this case, under appropriate
conditions it is possible to obtain \emph{trihamiltonian} Darboux-Nijenhuis
coordinates, having the most simple and natural property that one could
imagine in this context:

\begin{Prop}{}
Let \mbox{$\Nq, \Nr$} be the two tensors defined by (\ref{nij2}) from a
compatible trihamiltonian structure
on a $2m$--dimensional manifold. If
\begin{enumerate}
\item
all the eigenspaces of $\Nq$ and $\Nr$ coincide (equivalently, $\Nq$ and
$\Nr$ have the same
centraliser);
\item
both $\Nq$ and $\Nr$ have $m$ distinct eigenvalues, forming together a set
of $2m$
independent functions, that we denote by $\lambda_i$ and
$\mu_i$ respectively $(i=1,\ldots,m)$;
\item for any pair of eigenvalues $\lambda_i$ and $\mu_i$, 
corresponding to the same common
eigenspace,  one has
\BD
{\Pois{\lambda_i}{\mu_i}}_{_{\Pp}}=1;
\ED
\end{enumerate}
then the eigenvalues of $\Nq$ and $\Nr$ (respectively denoted by
$\lambda_i$ and $\mu_i$) form a
Darboux-Nijenhuis coordinate system.

\begin{Dim}
For any Nijenhuis recursion operator $N$
and any of its eigenvaules $\lambda$, it is always
true (see \cite{Magri3}) that $N^* d\lambda = \lambda d\lambda$.
The fact that $\Nq$ has $m$ independent eigenvalues
implies that all its eigenspaces are
bidimensional, and the same holds for $\Nr$. We have also assumed that the
eigenvalues
of $\Nq$ and $\Nr$ corresponding to the $i$-th common eigenspace are 
independent functions.
Therefore, the eigenspace itself is
spanned by the two differentials $d\lambda_i$ and $d\mu_i$:
\BD
\begin{array}{ccc}
\Nq^* d\lambda_i = \lambda_i d\lambda_i, & \qquad &\Nr^* d\lambda_i = \mu_i
d\lambda_i, \\
\Nq^* d\mu_i = \lambda_i d\mu_i, & &\Nr^*  d\mu_i = \mu_i d\mu_i.
\end{array}
\ED
 From the definition (\ref{nij2}) and the trivial fact that $\Nq\Pp = \Q =
\Pp\Nq^*$, at any point where
$\lambda_j\not=0$ one has
\begin{eqnarray*}
{\Pois{\lambda_i}{\lambda_j}}_{_{\Pp}} &=& \frac{1}{\lambda_j}\Dual{\Pp
d\lambda_i}{\lambda_j d\lambda_j}
= \frac{1}{\lambda_j}\Dual{\Pp d\lambda_i}{\Nq^* d\lambda_j} \\
&=& \frac{1}{\lambda_j}\Dual{\Pp\Nq^* d\lambda_i}{d\lambda_j}
= \frac{\lambda_i}{\lambda_j}{\Pois{\lambda_i}{\lambda_j}}_{_{\Pp}}
\end{eqnarray*}
and, since $\lambda_i\not=\lambda_j$ for $i\not= j$, one should have
\mbox{${\Pois{\lambda_i}{\lambda_j}}_{_{\Pp}}=0$} for all $i,j$. In a
similar way, using the recursion tensor
$\Nr$ one obtains
\mbox{${\Pois{\mu_i}{\mu_j}}_{_{\Pp}}=0$} for all $i,j$. Furthermore,
\begin{eqnarray*}
{\Pois{\lambda_i}{\mu_j}}_{_{\Pp}} &=& \frac{1}{\lambda_j}\Dual{\Pp
d\lambda_i}{\lambda_j d\mu_j}
= \frac{1}{\lambda_j}\Dual{\Pp d\lambda_i}{\Nq^* d\mu_j} \\
&=& \frac{1}{\lambda_j}\Dual{\Pp\Nq^* d\lambda_i}{d\mu_j}
= \frac{\lambda_i}{\lambda_j}{\Pois{\lambda_i}{\mu_j}}_{_{\Pp}}
\end{eqnarray*}
which entails \mbox{${\Pois{\lambda_i}{\mu_j}}_{_{\Pp}}=0 $} for \mbox{$ i
\neq j$}. The above results extend
by continuity to points where $\lambda_j=0$. The additional 
normalisation condition (3) ensures that
\mbox{$(\lambda_i,\mu_i)$} are canonical coordinates for $\Pp$; by 
construction, both
tensors $\Nq$
and $\Nr$ are diagonal in these coordinates, which therefore are
Darboux-Nijenhuis for the trihamiltonian
structure.
\end{Dim}
\end{Prop}

In other terms, if two Poisson pencils are available, and the two 
recursion operators
are ``independent" and ``compatible" in the sense given above, then 
\emph{all} the Darboux--Nijenhuis
variables are obtained as eigenvalues (in the usual bihamiltonian 
setup, only \emph{half}
of these variables are defined as eigenvalues). The theorem also 
clarifies that trihamiltonian
structures generated by single recursion operator, $\Q=N\Pp$ and 
$\R=N\Q=N^2\Pp$, sometimes
encountered in the literature, are not suitable for our purposes.

To obtain a twofold Poisson--Nijenhuis manifold also
when the Poisson tensor $\Pp$ is degenerate, we exploit a projection technique,
which has been already used for a different purpose in \cite{Deg}.

Let us start with two compatible Poisson tensors $\Pp$ and $\Q$.
The rank of $\Pp$ is assumed for simplicity
to be constant on the phase manifold
$\mathcal{M}$ (in the sequel, \mbox{$\dim\mathcal{M}=2m+k$}, $k$ 
being the corank of $\Pp$).
The kernel of $\Pp$ is then pointwise spanned by the
differentials of $k$ Casimir functions $c^\alpha$. The main 
ingredient in our approach is
a set of $k$ independent vectorfields $Z_\alpha$ (spanning a distribution
$\mathcal{Z}$) having these properties:
\BE\label{propcampi}
\begin{array}{rll}
i) & \mbox{normalization:} & Z_\alpha(c^\beta)=\delta_\alpha^\beta; \\
ii) & \mbox{integrability:} & [Z_\alpha,Z_\beta] \in \mathcal{Z}; \\
iii) & \mbox{symmetry for $\Pp$:} & \mathcal{L}_{Z_\alpha}\Pp=0.
\end{array}
\EE
The normalization property {\it (i)} entails that all the 
vectorfields $Z_\alpha$ are transversal
to the symplectic leaves of $\Pp$. The concrete possibility of 
finding such a set of transversal
vectorfields is not ensured {\it a priori}. Here we shall assume 
their existence, but in the last
section of the article we will give an explicit recipe to find them 
for a relevant class of
trihamiltonian structures.

The vectorfields
$Z_\alpha$ induce a decomposition on
$T\mathcal{M}$ and
$T^\ast\mathcal{M}$:
\begin{eqnarray*}
\forall \theta \in T^\ast\mathcal{M}
&&
\left\{\begin{array}{l}
\theta_\perp = \Dual{Z_\alpha}{\theta} \D c^\alpha \\
\theta_{/\!/} = \theta - \theta_\perp
\end{array}\right. \\
\forall X \in T\mathcal{M}
&&
\left\{\begin{array}{l}
X_\perp = X(c^\alpha)Z_\alpha \\
X_{/\!/} = X - X_\perp
\end{array}\right.
\end{eqnarray*}

\begin{Lemma}
The decomposition satisfies the following properties:
\begin{eqnarray*}
\Dual{\theta_{/\!/}}{Z_\alpha} &=& 0, \\
X_{/\!/}(c^\alpha) &=& 0.
\end{eqnarray*}
\begin{Dim}
$\Dual{\theta_{/\!/}}{Z_\alpha} = \Dual{\theta - \theta_\perp}{Z_\alpha}
= \Dual{\theta}{Z_\alpha} - \Dual{\theta}{Z_\beta}Z_\alpha(c^\beta) = 0$; \\
$X_{/\!/}(c^\alpha) = X(c^\alpha) - X_\perp(c^\alpha) = X(c^\alpha) - 
X(c^\beta)Z_\beta(c^\alpha)
= 0$.
\end{Dim}
\end{Lemma}

\noindent Using the above--defined decomposition, one proves that

\begin{Lemma}
The assumptions (\ref{propcampi}) imply $[Z_\alpha,Z_\beta]=0$.
\begin{Dim}
\mbox{$[Z_\alpha,Z_\beta] \in \mathcal{Z}$} entails
\mbox{$[Z_\alpha,Z_\beta]_{/\!/}=0$}, but one also has 
\mbox{$[Z_\alpha,Z_\beta]_\perp=0$}
because $[Z_\alpha,Z_\beta](c^\gamma) = Z_\alpha(\delta_\beta^\gamma) -
Z_\beta(\delta_\alpha^\gamma) = 0$.
\end{Dim}
\end{Lemma}

We can introduce a new tensor $\tilde\Q$ by setting
\BE\label{defor}
\tilde\Q(\theta,\phi) = \Q(\theta_{/\!/},\phi_{/\!/});
\EE
the kernel of the ``deformed'' tensor $\tilde\Q$ contains all the 
differentials of the
Casimir functions of $\Pp$. Still,  
the tensor $\tilde\Q$ is not
automatically a Poisson tensor. 

From the beginning, we have
assumed that $\mathcal{L}_{Z_\alpha}\Pp = 0$ (\ref{propcampi}). This is equivalent to
requiring that $\Pp$ be projectable along the flows of the
vectorfields $Z_\alpha$, i.e.~for any pair of functions $(f, g)$ such that
$Z_\alpha(f)=Z_\alpha(g)=0$ everywhere, their Poisson bracket should also be
constant along the same
flows: $Z_\alpha({\Pois{f}{g}}_{\Pp})=0$. 

So far, nothing ensures that the new tensor
$\tilde\Q$ is projectable in the same sense. Whenerver it turns out to
be so, i.e.
\BE\label{proj1}
\mathcal{L}_{Z_\alpha}\tilde\Q = 0,
\EE
then the $\tilde\Q$--bracket $\tilde\Q(df,dg)$ can be reduced to
$Z_\alpha$--invariant functions as well. For a $Z_\alpha$--invariant 
function $f$ one has \mbox{$df
= (df)_{/\!/}$}, so the $\tilde\Q$--bracket of such functions 
coincide with their
$\Q$--bracket. Then, if $\tilde\Q$ is projectable, the Jacobi
identity is straightfowardly proved for $Z_\alpha$--invariant functions. We claim that
(\ref{proj1}), supplemented by some auxiliary conditions, ensures both the
Jacobi identity and the compatibility with $\Pp$ for the $\tilde\Q$--bracket
of arbitrary functions on $\mathcal{M}$.

\begin{Prop}{}
If (i) all the vectorfields $Z_\alpha$ fulfill both (\ref{propcampi}) and
(\ref{proj1});\\
(ii) all the Casimir functions $c^\alpha$ are in mutual involution 
with respect to $\Q$, and
\\
(iii) the functions $c^\alpha$ generate bihamiltonian vectorfields, 
i.e.~one can find $k$
functions
$h^\alpha$ such that \mbox{$V^\alpha
\equiv \Q\D c^\alpha = \Pp\D h^\alpha$},\\
then $\tilde\Q$ is a Poisson tensor compatible with $\Pp$.
\end{Prop}\medskip

The proof requires several steps.

\begin{Lemma}{}
If the vectorfields $V^\alpha\equiv \Q\D c^\alpha$ are hamiltonian 
also with respect to $\Pp$,
then the Poisson tensor $\Q$ and the tensor $\tilde\Q$ differ by a 
Lie derivative of
$\Pp$:
\BE\label{Qdef}
\tilde\Q = \Q + \mathcal{L}_{X_{\Q}}\Pp,
\EE
where (summation is understood on repeated indices)
\BE\label{defvec}
X_{\Q} = h^\alpha Z_\alpha
\EE
\begin{Dim}
One has
\begin{eqnarray*}
\tilde\Q(df,dg) &=& \Q((df)_{/\!/}, (dg)_{/\!/}) =
\Q(df - (df)_\perp, dg - (dg)_\perp) \\
&=& \Q(df,dg) + Z_\alpha(f)Z_\beta(g)\Q(dc^\alpha,dc^\beta) - \\
&& - Z_\alpha(g)Z_\beta(g)\Q(df,dc^\alpha) - Z_\alpha(f)\Q(dc^\alpha,dg) \\
&=& \Q(df, dg) -
Z_\alpha(g)Z_\beta(g)\Pp(df,dh^\alpha) - Z_\alpha(f)\Pp(dh^\alpha,dg),
\end{eqnarray*}
which proves the statement since
\begin{eqnarray*}
(\mathcal{L}_{X_{\Q}}\Pp)(df, dg) &=&
h^\alpha\left[Z_\alpha({\Pois{f}{g}}_{\Pp}) -
{\Pois{Z_\alpha(f)}{g}}_{\Pp} - {\Pois{f}{Z_\alpha(g)}}_{\Pp}\right] - \\
&& - Z_\alpha(f){\Pois{h^\alpha}{g}}_{\Pp} -
Z_\alpha(g){\Pois{f}{h^\alpha}}_{\Pp} \\
&=& h^\alpha\left[(\mathcal{L}_{Z_\alpha}\Pp)(df,dg)\right] -
Z_\alpha(f){\Pois{h^\alpha}{g}}_{\Pp} -
Z_\alpha(g){\Pois{f}{h^\alpha}}_{\Pp} \\
&=& - Z_\alpha(f){\Pois{h^\alpha}{g}}_{\Pp} -
Z_\alpha(g){\Pois{f}{h^\alpha}}_{\Pp}.
\end{eqnarray*}
This also proves that, under the given hypotheses, 
\mbox{$\mathcal{L}_{X_{\Q}}\Pp =
V^\alpha \wedge Z_\alpha$}.
\end{Dim}
\end{Lemma}

If condition (iii) in the statement of Prop.(3.6) is 
fulfilled, then (\ref{proj1})
can be rewritten in terms of the tensor $\Q$, using the previous Lemma:

\begin{Lemma}{}
Upon assuming (\ref{Qdef}), (\ref{proj1}) is equivalent to
\BE\label{proj2}
\mathcal{L}_{Z_\alpha}\Q = [V^\beta,Z_\alpha] \wedge Z_\beta
\EE
\begin{Dim} Here and in the sequel we exploit the properties of the 
Schouten bracket, a
bilinear operator on contravariant, antisymmetric tensors of 
arbitrary rank $p$ ($p$-vectors). We
refer the reader to [Vaisman] for the general theory; the properties that we
shall use are the following:
\begin{itemize}
\item if $P$ is a $p$-vector and $Q$ is a $q$-vector, then
$\Schouten{P}{Q}=(-1)^{pq}\Schouten{Q}{P}$;
\item if, moreover, $R$ is a $r$-vector,
$\Schouten{P}{Q\wedge R}=\Schouten{P}{Q}\wedge R+(-1)^{q(p+1)}Q\wedge 
\Schouten{P}{R}$;
\item if $X$ is an ordinary vectorfield, then 
$\Schouten{X}{P}\equiv\mathcal{L}_{X}P$;
\item
$\mathcal{L}_{X}\Schouten{P}{Q}=\Schouten{\mathcal{L}_{X}P}{Q}+\Schouten{P}{\mathcal{L}_{X}Q}$;
\item if $P$ is a bivector, then it is a Poisson tensor if and only if
$\Schouten{P}{P}=0$;
\item if $P$ and $Q$ are both Poisson tensors, then they are 
compatible if and only if
$\Schouten{P}{Q}=0$.
\end{itemize}
Proving the Lemma is now straightforward:
\begin{eqnarray*}
\mathcal{L}_{Z_\alpha}\Q &=& \mathcal{L}_{Z_\alpha}(\tilde\Q - 
\mathcal{L}_{X_{\Q}}\Pp) \\
&=& -\mathcal{L}_{Z_\alpha}\mathcal{L}_{X_{\Q}}\Pp \\
&=& -\Schouten{Z_\alpha}{V^\beta \wedge Z_\beta} \\
&=& -[Z_\alpha,V^\beta] \wedge Z_\beta -V^\beta \wedge [Z_\alpha,Z_\beta] \\
&=& [V^\beta,Z_\alpha] \wedge Z_\beta
\end{eqnarray*}
\end{Dim}
\end{Lemma}

The condition (\ref{proj2}) is also a projectability condition:
it means that the tensor $\Q$ itself can be reduced to 
$Z_\alpha$--invariant functions, but in
this case one {\it cannot} identify the image of $\Q$ through 
projection (on the quotient space)
with the {\it restriction} of $\Q$ on a submanifold transversal to 
the vectorfields $Z_\alpha$
(for instance, a symplectic leaf of $\Pp$). As a matter of fact, we 
have seen that the projection
of $\Q$ would instead coincide with the restriction of $\tilde\Q$.

Now we can eventually prove the main statement.

\begin{Dim} We need to show that $\Schouten{\tilde\Q}{\tilde\Q}=0$ and
$\Schouten{\Pp}{\tilde\Q}=0$. The second equality is easy to prove. 
By assumption,
\mbox{$\Schouten{\Pp}{\Pp}=\Schouten{\Q}{\Q}=\Schouten{\Pp}{\Q}=0$}. 
Taking the Lie derivative of
$\Schouten{\Pp}{\Pp}$ one finds that
$\Schouten{\Pp}{\mathcal{L}_X\Pp}=0$ for any vectorfield $X$,
and from (\ref{Qdef}) it follows that
\mbox{$\Schouten{\Pp}{\tilde\Q}=\Schouten{\Pp}{\Q +
\mathcal{L}_{X_{\Q}}\Pp}\equiv 0$}. To prove the first
equality, observe furthemore that the vectorfields
$V^\alpha$ are assumed to be hamiltonian for $\Q$, hence 
\mbox{$\Schouten{\Q}{V^\alpha}
\equiv \mathcal{L}_{V^\alpha}\Q=0$}. Then
\begin{eqnarray*}
\Schouten{\tilde\Q}{\tilde\Q} &=& \Schouten{\Q}{\Q} + 
2\Schouten{\Q}{V^\alpha \wedge
Z_\alpha} + \Schouten{\mathcal{L}_{X_{\Q}}\Pp}{V^\alpha \wedge Z_\alpha} \\
&=& 2\Schouten{\Q}{V^\alpha} \wedge Z_\alpha -
2V^\alpha \wedge \Schouten{\Q}{Z_\alpha} + \\
&& \Schouten{\mathcal{L}_{X_{\Q}}\Pp}{V^\alpha} \wedge Z_\alpha -
V^\alpha \wedge \Schouten{\mathcal{L}_{X_{\Q}}\Pp}{Z_\alpha} \\
&=& 2V^\alpha \wedge [Z_\alpha,V^\beta] \wedge Z_\beta +
\Schouten{V^\alpha}{V^\beta \wedge Z_\beta} \wedge Z_\alpha -
V^\alpha \wedge \Schouten{Z_\alpha}{V^\beta \wedge Z_\beta} \\
&=& 2V^\alpha \wedge [Z_\alpha,V^\beta] \wedge Z_\beta +
[V^\alpha,V^\beta] \wedge Z_\beta \wedge Z_\alpha +
V^\beta \wedge [V^\alpha,Z_\beta] \wedge Z_\alpha - \\
&& V^\alpha \wedge [Z_\alpha,V^\beta] \wedge Z_\beta -
V^\alpha \wedge V^\beta \wedge [Z_\alpha,Z_\beta] \\
&=& 2V^\alpha \wedge [Z_\alpha,V^\beta] \wedge Z_\beta -
V^\alpha \wedge [Z_\alpha,V^\beta] \wedge Z_\beta -
V^\alpha \wedge [Z_\alpha,V^\beta] \wedge Z_\beta \\
&=& 0
\end{eqnarray*}
\end{Dim}

We have seen that $\Pp$ and $\tilde\Q$ are compatible and can both be 
reduced by projection along
the flows of the vectorfields $Z_\alpha$ on any symplectic leaf of 
$\Pp$. Then, on each leaf one
can define a recursion tensor and look for Darboux--Nijenhuis 
coordinates; but there is no evidence
that such coordinates can be extended to some neighborhood of the 
symplectic leaf in
$\mathcal{M}$. However, a Nijenhuis tensor
$\Nq $ can be directly defined on the full manifold $\mathcal{M}$ 
in the following way. To any
vectorfield $X$ over
$\mathcal{M}$ one can associate the vectorfield
$X_{/\!/}$ which is, by construction, everywhere tangent to the 
simplectic leaves of $\Pp$. Consider
now any one--form $\theta_X$ such that \mbox{$X_{/\!/}=\Pp\theta_X$}; we set
\BE\label{NdelQdef}
\Nq  X = \tilde\Q \theta_X.
\EE
The tensor $\Nq$ is actually independent of the choice of 
$\theta_X$, because the latter is
defined up to an element of the kernel of $\Pp$, which is also the 
kernel of $\tilde\Q$. From the
definition, it follows immediately that
\mbox{$\Nq  Z_\alpha = 0$}, and that for any hamiltonian 
vectorfield \mbox{$X_f=\Pp df$} one has
\mbox{$\Nq X_f=\tilde\Q df$}.

\begin{Prop}{}
Under the same hypotheses of the previous proposition, the tensor 
$\Nq $ defined by
(\ref{NdelQdef}) is a Nijenhuis tensor.
\begin{Dim}
We need to show that the Nijenhuis torsion tensor $T_N$ of $\Nq $ vanishes:
\BD
T_N(X,Y)\equiv[\Nq X,\Nq Y]-\Nq [\Nq X,Y]-\Nq [X,\Nq Y]+\Nq ^2[X,Y] 
= 0
\ED
The Nijenhuis torsion tensor acts pointwise on vectors; thus, it is 
sufficient to show
that $T_N(X,Y)$ vanishes on each pair of elements of a basis of the 
tangent space to $\mathcal{M}$
at any point. In our case, such a basis is provided by a set of 
$\Pp$--hamiltonian
vectorfields spanning the tangent spaces to the symplectic leaves, 
and by the transversal
vectorfields $Z_\alpha$. We already know that
\mbox{$\Nq Z_\alpha=0$} and
\mbox{$[Z_\alpha,Z_\beta]=0$}, thus \mbox{$T_N(Z_\alpha,Z_\beta)=0$} 
for any pair $(\alpha,\beta)$.
More generally, for any vectorfield $X$
\begin{eqnarray*}
T_N(X,Z_\alpha) &=& \Nq (\Nq [X,Z_\alpha]-[\Nq X,Z_\alpha]) \\
&=& \Nq (\mathcal{L}_{Z_\alpha}(\Nq X)-\Nq \mathcal{L}_{Z_\alpha}X) \\
&=& \Nq (\mathcal{L}_{Z_\alpha}(\Nq )\Pp\theta_X)
\end{eqnarray*}
but since 
\mbox{$\mathcal{L}_{Z_\alpha}\tilde{\Q}=\mathcal{L}_{Z_\alpha}\Pp=0$}, 
one has
\mbox{$\mathcal{L}_{Z_\alpha}(\Nq )\Pp=0$}.

Finally, we evaluate the torsion tensor on two hamiltonian 
vectorfields \mbox{$X_f=\Pp df$} and
\mbox{$X_g=\Pp dg$}; for notational convenience, we set \mbox{$Y_f=\Q 
df$}. This part of
the proof is well known and can be found, in greater detail, in 
[Marsico]: it is shown there that
the compatibility of $\tilde{\Q}$ and $\Pp$, i.e. $\Schouten{\Pp}{\tilde\Q}=0$,
implies
\BE\label{tiz1}
[X_f,Y_g]+[Y_f,X_g]=X_{{\Pois{f}{g}}_{\tilde{\Q}}}+Y_{{\Pois{f}{g}}_{\Pp}}.
\EE
Using the fact that \mbox{$Y_f=\Nq X_f$}, applying once again 
$\Nq $ to both sides of
(\ref{tiz1}), and rearranging terms, one gets
\BD
Y_{{\Pois{f}{g}}_{\tilde{\Q}}}=\Nq [X_f,\Nq X_g]+\Nq [\Nq X_f,X_g]-\Nq ^2[X_f,X_g];
\ED
on the oter hand,
\mbox{$Y_{{\Pois{f}{g}}_{\tilde{\Q}}}=[Y_f,Y_g]=[\Nq X_f,\Nq X_g]$} and so
\BD
T_N(X_f,X_g)=0.
\ED
\end{Dim}
\end{Prop}

These results suggest the following strategy:
given a trihamiltonian structure, choose one of the Poisson tensors,
say $\Pp$; a complete set of Casimir
functions
$c^\alpha$ can be directly read out from the ``fundamental molecule'', as
well as the functions $h^\alpha$ and
$k^\alpha$ used in (\ref{defvec}). In this case the vectorfields $\Q d
c^\alpha$ and $\R d c^\alpha$ are automatically commuting, so the conditions {\it
(ii)} and {\it (iii)} of Prop.(3.6) are fulfilled. If one is able to find a 
complete set of vectorfields fulfilling all of (\ref{propcampi}) and
being symmetries of both $\tilde\Q$ and $\tilde\R$ (in the
sequel, we shortly write {\it ``a good set of transversal symmetries''}), then one can
apply the above  procedure to each of the pairs $(\Pp,\Q)$ and $(\Pp,\R)$ separately,
obtaining in this way a pair of Nijenhuis tensors $\Nq $ and $\Nr $ on the full
manifold $\mathcal{M}$.

This fact, however, does not yet ensure the existence of 
Darboux--Nijenhuis coordinates. The proof of the existence of Darboux--Nijenhuis 
coordinates has been given in Prop.(3.3) only for pairs of Nijenhuis 
tensors fulfilling some additional requirements, first of all that of being 
non--degenerate. We have not addressed the 
general problem of the existence
of a canonical form for a non--regular Nijenhuis tensor; incidentally, to our
knowledge non--regular Nijenhuis tensors have never been previously considered in
connection  with finite--dimensionl
integrable systems. In the 
next section, we will rather characterise
a particular class of trihamiltonian systems for which 
Darboux--Nijenhuis coordinates exist, are separation variables for 
the Hamiltonians occurring in
the ``fundamental molecule'', and can even be constructed without
having to compute eigenvalues.
Although the assumptions that we
make could seem rather artificial and difficult to test in practice, 
we will eventually show that
this class of systems is not empty, and contains relevant examples 
described by Lax equations (with
spectral parameter).

\subsection{Sklyanin separation of trihamiltonian systems}

We first recall the separability criterion introduced by Sklyanin
\cite{Skly}. Let $h_i$ be $m$
hamiltonian in involution for a \emph{nondegenerate} Poisson tensor $P$,
and let
\mbox{$(\lambda_i,\mu_i)_{i=1,\ldots,m}$} a system of \emph{canonical}
coordinates for $P$. If $m$
functions $W_i$ of $m+2$ variables exist such that
\begin{eqnarray}
W_1\left(\lambda_1,\mu_1;h_1(\lambda_i,\mu_i),\ldots,h_m(\lambda_i,\mu_i)\right)
& = &
0 \nonumber\\
\vdots\qquad\\ \label{separ}
W_m\left(\lambda_m,\mu_m;h_1(\lambda_i,\mu_i),\ldots,h_m(\lambda_i,\mu_i)\right)
& = &
0 \nonumber
\end{eqnarray}
identically, then all
hamiltonians $h_i$ are separable in the coordinates \mbox{$(\lambda_i,\mu_i)$}.

This setting refers to the symplectic case (as it should be, since the notion of
separability of the Hamilton--Jacobi equation makes sense only in symplectic
manifolds,  namely on cotangent bundles). In the case of degenerate Poisson manifolds,
one can discuss separability on symplectic leaves. Upon reduction, one may expect to
find Sklyanin functions $W_i$ depending on auxiliary parameters labeling the
symplectic leaves (in our case, the Casimir functions $c^\alpha$). This is indeed
what happens in the example that we shall now discuss: first, we introduce a set of
equations built using the common Casimir function $f_{\lambda\mu}$ (generating the
``fundamental molecule'') and the transversal vectorfields $Z_\alpha$, and we list
sufficient conditions for the roots of these equations to form, together with the
Casimir functions $c^\alpha)$, a system of generalised Darboux--Nijenhuis coordinates
$(\lambda_i,\mu_i,c^\alpha)$ on $\mathcal{M}$. Then, we produce $m$ functions
$W_i$, each depending on the $i$-th pair of coordinates $(\lambda_i,\mu_i)$, on all
the $k$ Casimir functions $c^\alpha$, and on other $m$ arguments. Once the latter are
replaced by the remaining $m$ Hamiltonians $h^a_b(\lambda_j,\mu_j,c^\alpha)$ of the
fundamental molecule, the functions $W_i$ vanish identically for any value
of the coordinates $c^\alpha$, thus on all symplectic leaves simultaneously.

As previously, we assume
that all the three Poison tensors
$(\Pp,\Q,\R)$ have the same rank $2m$, and the dimension of the manifold is
$2m+k$. We also assume that the
common Casimir function $f_{\lambda\mu}$ is \emph{complete}, i.e.~that
among its coefficients one can
find $m+k$ independent functions, including $k$ Casimir functions for each
of the three Poisson tensors
(different tensors may indeed have some common Casimir functions). In the
cases that we shall consider (for instance, the
trihamiltonian spaces $\gl(r)^n$), the polynomial $f_{\lambda\mu}$ has
\emph{exactly} $m+k$ non--constant
coefficients.

The leading role in our construction is played by the
derivatives of the common Casimir function
$f_{\lambda\mu}$ along the transversal vectorfields $Z_\alpha$. We denote
these $k$ functions (still
depending on the two parameters
$\lambda,\mu$) by
\BE\label{skly}
S_\alpha(\lambda,\mu) = Z_\alpha(f_{\lambda\mu}).
\EE
The letter $S$ is chosen because of the coincidence with Sklyanin's
minors \cite{Skly}, in the particular case discussed in section (4).

Having assumed that the common Casimir function $f_{\lambda\mu}$ is
polynomial in
both parameters, the functions $S_\alpha(\lambda,\mu)$ are
polynomials as well. We shall prove that, whenever appropriate
conditions are verified, the
common roots $(\lambda_i,\mu_i)$ of the polynomials $S_\alpha(\lambda,\mu)$
are Darboux--Nijenhuis
coordinates and fulfill Sklyanin's separability condition (\ref{separ}).
Note that we use the abbreviated notation $F(\lambda\mu)|_{\lambda_i,\mu_i}$ for
$F(\lambda\mu)|_{\lambda=\lambda_i,\mu=\mu_i}$.

\begin{Prop}{}
If a good set of transversal symmetries $Z_\alpha$ fulfills in addition the
following requirements:
\begin{enumerate}
\item all second directional derivatives of the complete
common Casimir function $f_{\lambda\mu}$ vanish identically, i.e.
\BE
Z_\alpha(Z_\beta(f_{\lambda\mu}))=0 \qquad\mathrm{for\ all}\ \alpha,\beta;
\EE
\item the polynomials $S_\alpha(\lambda,\mu)=Z_\alpha(f_{\lambda\mu})$ have
$2m$ functionally independent
common roots \mbox{$\{\lambda_i,\mu_i\}$};
\item the following equality holds, and for any $i$ both sides are not
identically
vanishing for at least one pair
$(\alpha,\beta)$:
\end{enumerate}
\BE\label{nodegenere}
\left.{\Pois{S_\alpha(\lambda,\mu)}{S_\beta(\lambda,\mu)}}_{_{\Pp}}\right|_{\lambda_i,\mu_i}=
\left.\left(\frac{\partial S_\alpha}{\partial \lambda} \frac{\partial
S_\beta}{\partial \mu} -
\frac{\partial S_\alpha}{\partial \mu} \frac{\partial S_\beta}{\partial
\lambda}\right)\right|_{\lambda_i,\mu_i};
\EE
then, the $2m$ functions $(\lambda_i,\mu_i)$ form a system of
Darboux-Nijenhuis coordinates on each symplectic leaf of $\Pp$, and moreover
\mbox{$Z_\alpha(\lambda_i)=Z_\alpha(\mu_i)=0$}. 
\begin{Dim}
First, we compute the parameter--dependent vectorfield associated to the
common Casimir
function
$f_{\lambda\mu}$ under the deformed Poisson pencil $(\tilde\Q-\lambda\Pp)$:
\BE\label{formuletta}
(\tilde\Q-\lambda\Pp)\,df_{\lambda\mu} =
(\Q+\mathcal{L}_{X_{\Q}}\Pp-\lambda \Pp)\,df_{\lambda\mu} =
(\mathcal{L}_{X_{\Q}}\Pp)\,df_{\lambda\mu}.
\EE
But
\BE\label{laformula}
(\mathcal{L}_{X_{\Q}}\Pp)\,df_{\lambda\mu} = -\sum_{\alpha=1}^k
S_\alpha(\lambda,\mu) \cdot \Pp dh^\alpha:
\EE
in fact, from the definition of the
deformation vectorfields (\ref{defvec}), one sees that the vectorfield
$(\mathcal{L}_{X_{\Q}}\Pp)\,df_{\lambda\mu}$ acts on an arbitrary function
$g$ as follows:
\begin{eqnarray*}
\left[(\mathcal{L}_{X_{\Q}}\Pp)\,df_{\lambda\mu}\right](g) &=& -
\sum_{\alpha=1}^k Z_\alpha(f_{\lambda\mu}){\Pois{h^\alpha}{g}}_{_{\Pp}} -
\sum_{\alpha=1}^k Z_\alpha(g){\Pois{f_{\lambda\mu}}{h^\alpha}}_{_{\Pp}} \\
&=& -\sum_{\alpha=1}^k S_\alpha(\lambda,\mu) \cdot (\Pp dh^\alpha) (g).
\end{eqnarray*}
Taking the derivative along $Z_\alpha$ of (\ref{formuletta}), one obtains
\BD
(\tilde\Q-\lambda\Pp)\,dS_\alpha(\lambda,\mu) = -\sum_{\beta=1}^k
S_\beta(\lambda,\mu)\cdot \Pp
d[Z_\alpha(h^\beta)],
\ED
(we used the fact that $Z_\alpha$ is a symmetry for both $\Pp$ and $\tilde\Q$).

Since $(\lambda_i,\mu_i)$ is a pair of roots of $S_\alpha$, one finds that
$\left.[(\tilde\Q-\lambda\Pp)\,dS_\alpha(\lambda,\mu)]\right|_{\lambda_i,\mu_i}=0$
and that
\begin{eqnarray*}
0 = d[S_\alpha(\lambda_i,\mu_i)] &=&
[dS_\alpha(\lambda,\mu)]\Big|_{\lambda_i,\mu_i} \\ && +
\left.\frac{\partial S_\alpha}{\partial
\lambda}\right|_{\lambda_i,\mu_i}d\lambda_i
+\left.\frac{\partial S_\alpha}{\partial
\mu}\right|_{\lambda_i,\mu_i}d\mu_i
;
\end{eqnarray*}
then,
\BD
\left.\frac{\partial S_\alpha}{\partial \lambda}\right|_{\lambda_i,\mu_i}
(\tilde\Q d\lambda_i - \lambda_i \Pp d\lambda_i) + \left.\frac{\partial
S_\alpha}{\partial \mu}\right|_{\lambda_i,\mu_i} (\tilde\Q d\mu_i - \lambda_i \Pp d\mu_i)
= 0
\ED

On the other hand, 
\begin{eqnarray*}
Z_\beta(\lambda_i)\left.\frac{\partial S_\alpha}{\partial
\lambda}\right|_{\lambda_i,\mu_i}
+Z_\beta(\mu_i)\left.\frac{\partial S_\alpha}{\partial
\mu}\right|_{\lambda_i,\mu_i} = Z_\beta[S_\alpha(\lambda_i,\mu_i)] -
Z_\beta[S_\alpha(\lambda,\mu)]\Big|_{\lambda_i,\mu_i} = 0,
\end{eqnarray*}
having assumed $Z_\beta[S_\alpha(\lambda,\mu)]=0$ and
$S_\alpha(\lambda_i,\mu_i)=0$.
Let $\mathbf{S}$ be the $k \times 2$ matrix whose rows are
\mbox{$(\frac{\partial S_\alpha}{\partial
\lambda}, \frac{\partial S_\alpha}{\partial \mu})$}, and let $\mathbf{S}_{(i)}$
denote the same matrix after the substitution
$(\lambda,\mu)\rightarrow(\lambda_i,\mu_i)$. The condition
(\ref{nodegenere}) entails that $\mathbf{S}_{(i)}$
has maximal rank for all $i=1,\ldots,m$ so the previous results imply that
\begin{eqnarray*}
Z_\alpha(\lambda_i) &=& Z_\alpha(\mu_i) = 0 \\
\tilde\Q d\lambda_i &=& \lambda_i \Pp d\lambda_i \\
\tilde\Q d\mu_i &=& \lambda_i \Pp d\mu_i,
\end{eqnarray*}
and the same holds for $\tilde\R$. Repeating now the same argument used in the proof of
Prop.(3.3) one obtains that ${\Pois{\lambda_i}{\lambda_j}}_{_{\Pp}} = 0$ and
${\Pois{\mu_i}{\mu_j}}_{_{\Pp}} = 0$ for all $(i,j)$, and that
${\Pois{\lambda_i}{\mu_j}}_{_{\Pp}} = 0$ for $i \ne j$. 

To get the remaining canonical
bracket, one should instead rely on (\ref{nodegenere}): for any pair of polynomials
$F(\lambda,\mu)$ and $G(\lambda,\mu)$ such that
$F(\lambda_i,\mu_i)=G(\lambda_i,\mu_i)=0$, one has
\BD
\left.{\Pois{F(\lambda,\mu)}{G(\lambda,\mu)}}_{_{\Pp}}\right|_{\lambda_i,\mu_i}=
{\Pois{\lambda_i}{\mu_i}}_{_{\Pp}}\left.\left(\frac{\partial F}{\partial
\lambda} \frac{\partial
G}{\partial \mu} -
\frac{\partial F}{\partial \mu} \frac{\partial G}{\partial
\lambda}\right)\right|_{\lambda_i,\mu_i};
\ED
then, if (\ref{nodegenere}) holds and both sides are nonvanishing, one
concludes
\BD
{\Pois{\lambda_i}{\mu_i}}_{_{\Pp}}=1.
\ED
\end{Dim}
\end{Prop}

Next, we prove Sklyanin's separability
condition for the variables constructed according to the previous proposition.

\begin{Prop}{}
Let $(\lambda_i,\mu_i,c^\alpha)_{i=1,\ldots,m, \alpha=1,\ldots,k}$ be the
generalised Darboux-Nijenhuis coordinates associated to the projection
along the vectorfields $Z_\alpha$ according to Prop.(3.10). For each $i=1,\ldots,m$
\BE
f_{\lambda\mu}\big|_{\lambda_i,\mu_i}=p_i(\lambda_i,\mu_i)
\EE
where $p_i(\lambda,\mu)$ are polynomials with \emph{constant} coefficients.
Hence, the coordinates
$(\lambda_i,\mu_i)$ and the remaining $m$ hamiltonians $h^a_b$ (restricted
to the symplectic leaf
$\Sigma$) fulfill the separability condition (\ref{separ}), with
\BD
W_i(\lambda_i,\mu_i;h^a_b)=f_{\lambda\mu}\big|_{\lambda_i,\mu_i}-
p_i(\lambda_i,\mu_i)
\ED
\begin{Dim}
From the previous proposition, we know that
\BE\label{zero}
S_\alpha(\lambda_i,\mu_i)=0, \quad Z_\alpha(\lambda_i)=0, \quad
Z_\alpha(\mu_i)=0:
\EE
together with the normalization condition (\ref{propcampi}.$i$),
these imply that in the coordinate
system $(\lambda_i,\mu_i,c^\alpha)$ one has
\BE
Z_a\equiv\frac{\partial}{\partial c^\alpha}.
\EE
We denote the function obtained by replacing the spectral parameters
$(\lambda,\mu)$ with the pair of coordinates $(\lambda_i,\mu_i )$ by
$f_{(i)}\equiv f_{\lambda\mu}\big|_{\lambda_i,\mu_i}$. On
account of (\ref{zero}),
  \BD
Z_\alpha\left(f_{(i)}\right) = S_\alpha(\lambda_i,\mu_i) = 0
\ED
thus the $m$ functions $f_{(i)}$, which would in principle depend on
\emph{all} the coordinates $(\lambda_j,\mu_j,c^\alpha)$, actually do not
depend on the Casimir
coordinates $c^\alpha$. To prove that they depend only on the pair
$(\lambda_i,\mu_i)$, we exploit the properties of
Darboux--Nijenhuis coordinates:
for any function $g$, one has
\BD
{\Pois{g}{\lambda_j}}_{_{\Pp}}=\frac{\partial
g}{\partial\mu_j}\qquad\mathrm{and}\qquad
{\Pois{g}{\lambda_j}}_{_{\tilde\Q}}=\lambda_j\frac{\partial g}{\partial\mu_j}.
\ED
Thus, if for some $i$ one finds a function $g$ such that
$
{\Pois{g}{\lambda_j}}_{_{\tilde\Q}}-\lambda_i{\Pois{g}{\lambda_j}}_{_{\Pp}}=0
$
identically for any $j$, then $(\lambda_j-\lambda_i)
\frac{\partial
g}{\partial \mu_j}\equiv 0$ and the function $g$ cannot depend on $\mu_j$
for $j\not= i$.
If furthermore
$
{\Pois{g}{\mu_j}}_{_{\tilde\R}}-\mu_i{\Pois{g}{\mu_j}}_{_{\Pp}}
$ vanishes for any $j$ as well, then $g$ depends only on the pair
$(\lambda_i,\mu_i)$.
So what we need to prove is that, for any $j\not= i$,
\BE\label{quello}
{\Pois{f_{(i)}}{\lambda_j}}_{_{\tilde\Q}}-\lambda_i{\Pois{f_{(i)}}{\lambda_j}}_{
_{\Pp}}=
{\Pois{f_{(i)}}{\mu_j}}_{_{\tilde\R}}-\mu_i{\Pois{f_{(i)}}{\mu_j}}_{_{\Pp}}=
0.
\EE
The differential of the function $f_{(i)}$ is given
by
\BD
df_{(i)} = \left.\left(df_{\lambda\mu}\right)\right|_{\lambda_i,\mu_i}
+\left.\frac{\partial f_{\lambda\mu}}{\partial
\lambda}\right|_{\lambda_i,\mu_i}d\lambda_i+
\left.\frac{\partial f_{\lambda\mu}}{\partial
\mu}\right|_{\lambda_i,\mu_i}d\mu_i.
\ED
Thus, the vectorfield defined by applying the tensor
$(\tilde\Q-\lambda_i\Pp)$ to the differential of
the function $f_{(i)}$ can be obtained from (\ref{formuletta}) by replacing
the parameters
$(\lambda,\mu)$ with the $i$--th pair of coordinates $(\lambda_i,\mu_i)$, and
adding two terms proportional to $\frac{\partial f_{\lambda\mu}}{\partial
\lambda}$ and $\frac{\partial f_{\lambda\mu}}{\partial
\mu}$ respectively. Applying this vectorfield to a coordinate
$\lambda_j$, with $j\neq i$, one gets:
\begin{eqnarray*}
{\Pois{f_{(i)}}{\lambda_j}}_{_{\tilde\Q}}-\lambda_i{\Pois{f_{(i)}}{\lambda_j}}_{
_{\Pp}}
&=&
\left.{\Pois{f_{\lambda\mu}}{\lambda_j}}_{_{(\tilde\Q-\lambda\Pp)}}\right|_{\lambda_i,\mu_i}
+
\\ &&
\left({\Pois{\lambda_i}{\lambda_j}}_{_{\tilde\Q}}-\lambda_i{\Pois{\lambda_i}{\lambda_j}}_{_{\Pp}}
\right)\left.\frac{\partial f_{\lambda\mu}}{\partial
\lambda}\right|_{\lambda_i,\mu_i} +
\\ &&
\left({\Pois{\mu_i}{\lambda_j}}_{_{\tilde\Q}}-\lambda_i{\Pois{\mu_i}{\lambda_j}}
_{_{\Pp}}
\right)\left.\frac{\partial f_{\lambda\mu}}{\partial
\mu}\right|_{\lambda_i,\mu_i}.
\end{eqnarray*}
By hypothesis, $\lambda_j$ is in involution with both $\lambda_i$ and
$\mu_i$ for $j\neq i$; therefore, only
the first line survives, but due to
(\ref{formuletta}) the r.h.s.~is equal to
$-\sum_{\alpha=1}^k
S_\alpha(\lambda_i,\mu_i){\Pois{h^\alpha}{\lambda_j}}_{_{\Pp}}$, which once
again vanishes
on account of (\ref{zero}): half of (\ref{quello}) is proved.
Repeating the whole argument for the coordinate $\mu_i$, upon replacing
$\tilde\Q$ with
$\tilde\R$, one proves the full statement.
\end{Dim}
\end{Prop}

Thus, the search for separation variables is completely translated into the
problem of finding a set of ``good'' transversal vectorfields. We leave three questions
open. First, suppose that one were able to find (in some other way) a set of generalised
Darboux--Nijenhuis coordinates for the ``deformed'' Poisson triple
$(\Pp,\tilde\Q,\tilde\R)$; would these coordinates be roots of the polynomials $S_\alpha$
(without imposing further conditions)? 

Second question, is requirement (3) in Prop.(3.10) really necessary, or is it already
implied by the previous assumptions? We could not find any concrete example in which
(\ref{propcampi}) and the requirements (1), (2) of Prop.(3.10) are satisfied, but (3) fails
to hold. One could then suspect that the equality (3) can be derived from the other
(simpler) assumptions. The trouble with the requirement (3) is that it is both uneasy to
check and lacking a clear geometrical significance\footnote{In principle, it would be
possible to consider the manifold
$\mathcal{M}\times\Reali^2$ (with the spectral parameters $\lambda$, $\mu$ regarded
as two additional real coordinates), endowed with the direct
sum of the structure $\Pp$ on $\mathcal{M}$ with the Poisson structure on $\Reali^2$for
which $(\lambda,\mu)$ are canonical coordinates. Then, (3) could be rephrased by saying
that all the Poisson brackets of the functions $S_\alpha$ (regarded as functions of
$2m+k+2$ variables) should vanish on each submanifold described by the equations
$\lambda=\lambda_i$, $\mu=\mu_i$. Still, this does not seem to help very much.};
nevertheless, we have not yet succeeded in replacing it with another condition
equally ensuring that ${\Pois{\lambda_i}{\mu_i}}_{_{\Pp}}=1$.

Third open problem: under our assumptions, we have obtained Sklyanin's separation
condition with
$W_i=f_{(i)}-p_i(\lambda_i,\mu_i)$. In the
algebro--geometric setting, one deals with a more particular situation, namely
$p_i(\lambda_i,\mu_i)=p(\lambda_i,\mu_i)$ for a \emph{fixed} polynomial
$p(\lambda,\mu)$ not depending on
$i$, so all separation variables are (pairwise) roots of a single
polynomial $f_{\lambda\mu}-p(\lambda,\mu)$,
defining the \emph{spectral curve} of the system. At the moment, we do not
know which additional conditions
would ensure this stronger type of separability, which is encountered in
the examples that we discuss below. In the next subsection we shall
produce an example showing that
the occurrence of a single spectral curve does not follow
automatically from our assumptions.

\subsection{Canonical form of trihamiltonian structures in separation
coordinates}

As we have seen, given a trihamiltonian structure ($\Pp,\Q,\R$) and a
complete common Casimir polynomial
$f_{\lambda\mu}$, one needs to find a good set of transversal
vectorfields to produce separation
variables. Finding such vectorfields is, in general, a
difficult task.
On the other hand, upon assuming that such vectorfields exist, one can explicitly 
compute the components of all the relevant object as they would become in
separation coordinates, as we shall see now; in some cases, this provides a concrete
procedure to obtain ``backwards'' the change of variables.

From the previous discussion, we know that in a generalised Darboux--Nijenhuis coordinate
system:
\begin{enumerate}

\item as far as the \mbox{$2m \times 2m$} block corresponding 
to the coordinates
\mbox{$(\lambda_i,\mu_i)$} is considered, the tensor $\Pp$ is in canonical form, 
while the tensors $\tilde\Q$ and
$\tilde\R$ are obtained from $\Pp$ by applying the \emph{diagonal} 
recursion operators
$\Nq$ and $\Nr$, having the
coordinates $\lambda_i$ and $\mu_i$, respectively, as (double) eigenvalues;

\item the components of the complete tensors $\Pp$, $\tilde\Q$ and 
$\tilde\R$ in the
coordinates $(\lambda_i,\mu_i,c^\alpha)$ are obtained by simply 
adding $k$ null rows and
$k$ null columns to the respective $2m \times 2m$ matrix (by
``null'' we mean that all the corresponding entries are vanishing);

\item the transversal vectorfields $Z_\alpha$ are coordinate vectorfields:
$Z_\alpha=\frac{\partial}{\partial
c^\alpha}$;

\item the relation between the original Poisson tensors $\Q$ and $\R$ with
the ``deformed'' ones, $\tilde\Q$
and $\tilde\R$, is given by (\ref{Qdef});

\item each pair of conjugate coordinates $(\lambda_i,\mu_i)$ is a root of
the equation
$f_{\lambda\mu}-p_i(\lambda,\mu)=0$ for some polynomial $p_i$.
\end{enumerate}
The latter information allows one to find the explicit expression of all
the hamiltonians $h^i_j$ as functions
of
$(\lambda_i,\mu_i,c^\alpha)$, once fixed the
polynomials $p_i$ (which may be separately determined or arbitrarily
chosen, as we explain below). In fact,
let us assume as before that the common Casimir polynomial $f_{\lambda\mu}$
contains exactly $m+k$ independent
hamiltonians
$h^i_j$. Let us single out the $k$ hamiltonians which are Casimir functions
for $\Pp$, which we denote as
above by
$c^\alpha$, and denote the remaining (independent) hamiltonians as $h_A$,
with $A=1,\ldots,m$. We impose the
conditions
\begin{eqnarray}\label{hamils}
f_{(1)}(\lambda_1,\mu_1) &=& p_1(\lambda_1,\mu_1) \nonumber\\
&\vdots& \\
f_{(m)}(\lambda_m,\mu_m) &=& p_m(\lambda_m,\mu_m) \nonumber
\end{eqnarray}
which form a \emph{linear} system of $m$ independent equations in the
$m$ unknowns
$h_A$. Solving it, one finds $h_A=h_A(\lambda_i,\mu_i,c^\alpha)$.

Next, one produces the deformation vectorfields $X_{\Q}$ and $X_{\R}$
according to Lemma~(3.7). Then, one can
compute the components of the two Poisson tensors $\Q$ and $\R$:
\BE\label{PoisRic}
\Q = \tilde\Q - \mathcal{L}_{X_{\Q}}\Pp \qquad\mathrm{and}\qquad
\R = \tilde\R - \mathcal{L}_{X_{\R}}\Pp.
\EE
It is also important to remark that one can obtain as well the expression of
the polynomials
$S_\alpha(\lambda,\mu)=\frac{\partial f_{\lambda\mu}}{\partial c^\alpha}$.
This fact will be used in the
applications.

To fix the ideas, we work out a concrete example. We remark that in this way we shall
display a concrete case where the requirements of Prop.(3.10) can be directly tested,
showing that the set of systems fulfilling our requirements is indeed not empty (other
examples can be produced in the same way, starting from different ``fundamental
molecules''). Take the
$\gl(3)$ ``fundamental molecule'' represented in (\ref{cry1}). We set
$c^1\equiv h^0_2$, $c^2\equiv h^1_1$
and $c^3\equiv h^2_0$; we also simplify the notation for the remaining
hamiltonians by setting $h_1\equiv h^0_0$,
$h_2\equiv h^0_1$, and $h_3\equiv h^1_0$.
The common Casimir polynomial (apart from possible constant terms, which
may anyhow be
compensated in the
polynomials $p_i$) becomes
\BE\label{casimirobello}
f_{\lambda\mu}=h_1 + h_2\lambda + c_1\lambda^2 + h_3\mu + c_2\lambda\mu +
c_3\lambda\mu^2.
\EE
We leave the three polynomials $p_1$, $p_2$ and $p_3$ undetermined; 
for brevity, we write
$p_{(1)}$ for $p_1(\lambda_1,\mu_1)$, and so on.
Imposing (\ref{hamils}), the
hamiltonians read as follows:

\begin{eqnarray*}
h_1 = &\Big[&(\lambda_2\mu_3-\lambda_3\mu_2)p_{(1)} +
(\lambda_3\mu_1-\lambda_1\mu_3)p_{(2)} +
(\lambda_1\mu_2-\lambda_2\mu_1)p_{(3)} + \\
&& \left[ (\lambda_3-\lambda_2)\lambda_2\lambda_3\mu_1 +
(\lambda_1-\lambda_3)\lambda_1\lambda_3\mu_2 +
(\lambda_2-\lambda_1)\lambda_1\lambda_2\mu_3 \right] c_1 + \\
&& \left[ (\mu_3-\mu_2)\lambda_2\lambda_3\mu_1 +
(\mu_1-\mu_3)\lambda_1\lambda_3\mu_2 +
(\mu_2-\mu_1)\lambda_1\lambda_2\mu_3 \right] c_2 + \\
&& \left[ (\mu_2-\mu_3)\mu_2\mu_3\lambda_1 +
(\mu_3-\mu_1)\mu_1\mu_3\lambda_2 +
(\mu_1-\mu_2)\mu_1\mu_2\lambda_3 \right] c_3\ \Big] \cdot \\
&& \cdot \left( \lambda_1\mu_2-\lambda_2\mu_1 +
\lambda_2\mu_3-\lambda_3\mu_2 +
\lambda_3\mu_1-\lambda_1\mu_3 \right)^{-1}
\\ \\
h_2 = &\Big[&(\mu_2-\mu_3)p_{(1)} + (\mu_3-\mu_1)p_{(2)} +
(\mu_1-\mu_2)p_{(3)} + \\
&& \left[ ({\lambda_2}^2-{\lambda_3}^2)\mu_1 +
({\lambda_3}^2-{\lambda_1}^2)\mu_2 +
({\lambda_1}^2-{\lambda_2}^2)\mu_3 \right] c_1 + \\
&& \left[ (\lambda_2-\lambda_1)\mu_1\mu_2 +
(\lambda_1-\lambda_3)\mu_1\mu_3 +
(\lambda_3-\lambda_2)\mu_2\mu_3 \right] c_2 + \\
&& \left[ (\mu_2-\mu_1)\mu_1\mu_2 +
(\mu_1-\mu_3)\mu_1\mu_3 +
(\mu_3-\mu_2)\mu_2\mu_3 \right] c_3\ \Big] \cdot \\
&& \cdot \left( \lambda_1\mu_2-\lambda_2\mu_1 +
\lambda_2\mu_3-\lambda_3\mu_2 +
\lambda_3\mu_1-\lambda_1\mu_3 \right)^{-1}
\\ \\
h_3 = &\Big[&(\lambda_3-\lambda_2)p_{(1)} + (\lambda_3-\lambda_2)p_{(1)} +
(\lambda_3-\lambda_2)p_{(1)} + \\
&& \left[ (\lambda_1-\lambda_2)\lambda_1\lambda_2 +
(\lambda_3-\lambda_1)\lambda_1\lambda_3 +
(\lambda_2-\lambda_3)\lambda_2\lambda_3 \right] c_1 + \\
&& \left[ (\mu_1-\mu_2)\lambda_1\lambda_2 +
(\mu_3-\mu_1)\lambda_1\lambda_3 +
(\mu_2-\mu_3)\lambda_2\lambda_3 \right] c_2 + \\
&& \left[ ({\mu_3}^2-{\mu_2}^2)\lambda_1 +
({\mu_1}^2-{\mu_3}^2)\lambda_2 +
({\mu_2}^2-{\mu_1}^2)\lambda_3 \right] c_3\ \Big] \cdot \\
&& \cdot \left( \lambda_1\mu_2-\lambda_2\mu_1 +
\lambda_2\mu_3-\lambda_3\mu_2 +
\lambda_3\mu_1-\lambda_1\mu_3 \right)^{-1}
\end{eqnarray*}
The deformation vectorfields (\ref{defvec}) are
\BD
X_{\Q}=h_2\frac{\partial}{\partial
c^1}+h_3\frac{\partial}{\partial c^2},\qquad
X_{\R}=h_2\frac{\partial}{\partial
c^2}+h_3\frac{\partial}{\partial c^3}.
\ED
Starting from
\BD
\tiny \Pp = \left(
\begin{array}{ccccccccc}
0 & 0 & 0 & 1 & 0 & 0 & 0 & 0 & 0 \\
0 & 0 & 0 & 0 & 1 & 0 & 0 & 0 & 0 \\
0 & 0 & 0 & 0 & 0 & 1 & 0 & 0 & 0 \\
-1 & 0 & 0 & 0 & 0 & 0 & 0 & 0 & 0 \\
0 & -1 & 0 & 0 & 0 & 0 & 0 & 0 & 0 \\
0 & 0 & -1 & 0 & 0 & 0 & 0 & 0 & 0 \\
0 & 0 & 0 & 0 & 0 & 0 & 0 & 0 & 0 \\
0 & 0 & 0 & 0 & 0 & 0 & 0 & 0 & 0 \\
0 & 0 & 0 & 0 & 0 & 0 & 0 & 0 & 0
\end{array}
\right)
\ED
\BD
\tiny \tilde\Q = \left(
\begin{array}{ccccccccc}
0 & 0 & 0 & \lambda_1 & 0 & 0 & 0 & 0 & 0 \\
0 & 0 & 0 & 0 & \lambda_2 & 0 & 0 & 0 & 0 \\
0 & 0 & 0 & 0 & 0 & \lambda_3 & 0 & 0 & 0 \\
-\lambda_1 & 0 & 0 & 0 & 0 & 0 & 0 & 0 & 0 \\
0 & -\lambda_2 & 0 & 0 & 0 & 0 & 0 & 0 & 0 \\
0 & 0 & -\lambda_3 & 0 & 0 & 0 & 0 & 0 & 0 \\
0 & 0 & 0 & 0 & 0 & 0 & 0 & 0 & 0 \\
0 & 0 & 0 & 0 & 0 & 0 & 0 & 0 & 0 \\
0 & 0 & 0 & 0 & 0 & 0 & 0 & 0 & 0
\end{array}
\right)
  \qquad
\tilde\R = \left(
\begin{array}{ccccccccc}
0 & 0 & 0 & \mu_1 & 0 & 0 & 0 & 0 & 0 \\
0 & 0 & 0 & 0 & \mu_2 & 0 & 0 & 0 & 0 \\
0 & 0 & 0 & 0 & 0 & \mu_3 & 0 & 0 & 0 \\
-\mu_1 & 0 & 0 & 0 & 0 & 0 & 0 & 0 & 0 \\
0 & -\mu_2 & 0 & 0 & 0 & 0 & 0 & 0 & 0 \\
0 & 0 & -\mu_3 & 0 & 0 & 0 & 0 & 0 & 0 \\
0 & 0 & 0 & 0 & 0 & 0 & 0 & 0 & 0 \\
0 & 0 & 0 & 0 & 0 & 0 & 0 & 0 & 0 \\
0 & 0 & 0 & 0 & 0 & 0 & 0 & 0 & 0
\end{array}
\right)
\ED
one can obtain the matrix expressions for the tensors $\Q = \tilde\Q -
\mathcal{L}_{X_{\Q}}\Pp$ and $\R = \tilde\R - \mathcal{L}_{X_{\R}}\Pp$. They
coincide with the latter two matrices
above, respectively, as far as the $6\times 6$ upper left blocks are
concerned, while the remaining three
rows and three columns are rather complicated for both tensors (and it would be
pointless to write them down here). The fact that the tensors $\Pp$, $\tilde\Q$ and
$\tilde\R$ above are pairwise compatible Poisson tensors, as well as the fact that the
coordinate vecorfields $Z_\alpha=\frac{\partial}{\partial
c^\alpha}$ are symmetries of all of them, are trivially verified; then, is it enough to
reverse the steps of the proof of Prop.(3.6) to show that $\Q$ and
$\R$ are both Poisson tensors compatible with $\Pp$.
One can then check directly that the function
(\ref{casimirobello}) with the three hamiltonians computed above is a Casimir
function for $\Pp$, $\Q$ and $\R$, as expected. 

The polynomials 
$S_\alpha(\lambda,\mu;\lambda_i,\mu_i)$
are
\begin{eqnarray*}
S_1 &=& \frac{(\lambda_3-\lambda_2)\lambda_2\lambda_3\mu_1 +
(\lambda_1-\lambda_3)\lambda_1\lambda_3\mu_2 +
(\lambda_2-\lambda_1)\lambda_1\lambda_2\mu_3}
{\lambda_1\mu_2-\lambda_2\mu_1+\lambda_2\mu_3-\lambda_3\mu_2+\lambda_3\mu_1-\lambda_1\mu_3}
  + \\
&&  \left(\frac{({\lambda_2}^2-{\lambda_3}^2)\mu_1 +
({\lambda_3}^2-{\lambda_1}^2)\mu_2 +
({\lambda_1}^2-{\lambda_2}^2)\mu_3}
{\lambda_1\mu_2-\lambda_2\mu_1+\lambda_2\mu_3-\lambda_3\mu_2+\lambda_3\mu_1-\lambda_1\mu_3}
  \right)\lambda + \\
&&  \left(\frac{(\lambda_1-\lambda_2)\lambda_1\lambda_2 +
(\lambda_3-\lambda_1)\lambda_1\lambda_3 +
(\lambda_2-\lambda_3)\lambda_2\lambda_3}
{\lambda_1\mu_2-\lambda_2\mu_1+\lambda_2\mu_3-\lambda_3\mu_2+\lambda_3\mu_1-\lambda_1\mu_3}
  \right)\mu + \lambda^2 \\[10pt]
S_2 &=& \frac{(\mu_3-\mu_2)\lambda_2\lambda_3\mu_1 +
(\mu_1-\mu_3)\lambda_1\lambda_3\mu_2 + (\mu_2-\mu_1)\lambda_1\lambda_2\mu_3}
{\lambda_1\mu_2-\lambda_2\mu_1+\lambda_2\mu_3-\lambda_3\mu_2+\lambda_3\mu_1-\lambda_1\mu_3}
+ \\
&& \left( \frac{(\lambda_2-\lambda_1)\mu_1\mu_2 +
(\lambda_1-\lambda_3)\mu_1\mu_3 + (\lambda_3-\lambda_2)\mu_2\mu_3}
{\lambda_1\mu_2-\lambda_2\mu_1+\lambda_2\mu_3-\lambda_3\mu_2+\lambda_3\mu_1-\lambda_1\mu_3}
\right) \lambda + \\
&& \left( \frac{(\mu_1-\mu_2)\lambda_1\lambda_2 +
(\mu_3-\mu_1)\lambda_1\lambda_3 + (\mu_2-\mu_3)\lambda_2\lambda_3}
{\lambda_1\mu_2-\lambda_2\mu_1+\lambda_2\mu_3-\lambda_3\mu_2+\lambda_3\mu_1-\lambda_1\mu_3}
\right) \mu + \lambda\mu \\[10pt]
S_3 &=& \frac{(\mu_2-\mu_3)\mu_2\mu_3\lambda_1 +
(\mu_3-\mu_1)\mu_1\mu_3\lambda_2 + (\mu_1-\mu_2)\mu_1\mu_2\lambda_3}
{\lambda_1\mu_2-\lambda_2\mu_1+\lambda_2\mu_3-\lambda_3\mu_2+\lambda_3\mu_1-\lambda_1\mu_3}
+ \\
&& \left( \frac{(\mu_2-\mu_1)\mu_1\mu_2 +
(\mu_1-\mu_3)\mu_1\mu_3 + (\mu_3-\mu_2)\mu_2\mu_3}
{\lambda_1\mu_2-\lambda_2\mu_1+\lambda_2\mu_3-\lambda_3\mu_2+\lambda_3\mu_1-\lambda_1\mu_3}
\right) \lambda + \\
&& \left( \frac{({\mu_3}^2-{\mu_2}^2)\lambda_1 +
({\mu_1}^2-{\mu_3}^2)\lambda_2 + ({\mu_2}^2-{\mu_1}^2)\lambda_3}
{\lambda_1\mu_2-\lambda_2\mu_1+\lambda_2\mu_3-\lambda_3\mu_2+\lambda_3\mu_1-\lambda_1\mu_3}
\right) \mu + \mu^2
\end{eqnarray*}
This set of polynomials satisfies all the conditions of Prop.(3.10): first,
they do not depend any more
on the coordinates $c^\alpha$; second, the reader can check straightforwardly that 
$(\lambda_1,\mu_1)$,
$(\lambda_2,\mu_2)$ and
$(\lambda_3,\mu_3)$ are pairs of common roots of the polynomials $S_1$,
$S_2$ and $S_3$; as far as the third condition is concerned, one should check that
for any $i=1,2,3$ the two matrices
$\Big\|\Pois{S_\alpha}{S_\beta}_{\Pp}\Big\|_{\lambda_i,\mu_i}$ and
$\Big\|\frac{\partial S_{\alpha\phantom{\beta}}}{\partial\lambda}
\frac{\partial S_\beta}{\partial\mu}-
\frac{\partial S_{\alpha\phantom{\beta}}}{\partial\mu}
\frac{\partial S_\beta}{\partial\lambda}\Big\|_{\lambda_i,\mu_i}$ (with
$\alpha,\beta=1,2,3$) are not identically vanishing and coincide. Direct computation shows
that this is indeed the case: for $i=1$, the two matrices are both equal
to
\BD
\left(
\begin{array}{ccc}
0 & (\lambda_2\!-\!\lambda_1)(\lambda_3\!-\!\lambda_1) &
\begin{array}{c}
(\lambda_2\!-\!\lambda_1)(\mu_3\!-\!\mu_1)+\\+(\mu_2\!-\!\mu_1)(\lambda_3\!-\!\lambda_1)
\end{array} 
\\ \\
(\lambda_1\!-\!\lambda_2)(\lambda_3\!-\!\lambda_1) & 0 & 
(\mu_2\!-\!\mu_1)(\mu_3\!-\!\mu_1) \\ \\
\begin{array}{c}
(\lambda_1\!-\!\lambda_2)(\mu_3\!-\!\mu_1)+\\+(\mu_1\!-\!\mu_2)(\lambda_3\!-\!\lambda_1) 
\end{array}
&
(\mu_1\!-\!\mu_2)(\mu_3\!-\!\mu_1) & 0
\end{array}
\right) 
\ED
and similar expressions, with the appropriate permutiations of indices, are found for
$i=2,3$. 

It is worthwhile to remark that one can produce, choosing arbitrarily the polynomials
$p_i(\lambda,\mu)$, infinitely many families of hamiltonians which are separable according
to Sklyanin's criterion, but do not coincide with the coefficients of a single spectral
curve, unless one sets
$p_1(\lambda,\mu)\equiv p_2(\lambda,\mu)\equiv p_3(\lambda,\mu)\equiv p(\lambda,\mu)$.

Notice that the choice of the constant polynomials $p_i$ does not affect
the polynomials $S_\alpha$.
Actually, if one starts from a fixed trihamiltonian structure and is able
to find vectorfields $Z_\alpha$
fulfilling all the requirement listed in Prop.(3.10), then the polynomials
$p_i$ are determined \emph{a
posteriori} simply by plugging separately each pair of common roots of the
polynomials $S_\alpha$ into
$f_{\lambda\mu}$. The above reconstruction of
the trihamiltonian structure
goes in the reverse direction: the polynomials
$p_i$ are arbitrary and determine at the same time the hamiltonians and the
Poisson tensors $\Q$ and $\R$.

Thus, in our framework the constant terms in Sklyanin's separation polynomials
$W_i$, and \emph{a fortiori} in the
spectral curve (whenever it exists), are \emph{not} directly encoded in the
hamiltonian structure
underlying a dynamical system and its symmetries (some
aspects connected to the arbitrarity of the constant part of the the
spectral curve equation have been
addressed by J.~Harnad in \cite{Harnad}). We have seen that the 
constant polynomials
$p_i$ are
determined by the choice of a set of transversal vectorfields, or
equivalently of a system of separation
variables: possible different sets of polynomials $p_i$ -- and, eventually,
different spectral curves -- are
associated to different sets of separation variables. Indeed, from the
example given above one might infer
that different choices of the polynomials $p_i$ lead to different
hamiltonians (hence, to distinct dynamical
systems), but in fact these are -- by construction -- nothing but the same
hamiltonians expressed in two
different coordinate systems: the same holds for the components of both
$\Q$ and $\R$. The
transversal vectorfields, on the contrary, would have the same components
in both coordinate systems, but the
vectorfields themselves would not be the same, as happens for
the deformed structures $\tilde\Q$
and $\tilde\R$.

In conclusion, the spectral curve appears to be an \emph{additional datum}
with respect to the purely
hamiltonian structure of an integrable system; its hamiltonian
interpretation, however, is deeply connected
to the existence of particular canonical coordinates, as (separately)
suggested by Sklyanin and Magri. For trihamiltonian structures (with a
suitable projection onto a symplectic leaf), we have found a sound
connection between
separation coordinates and the vanishing of spectral polynomials.

\setcounter{equation}{0}
\section{Separation coordinates for Lax equations with
spectral parameter}

In this section we apply the techniques discussed so far to a particular class of dynamical
systems, represented by Lax equations with spectral parameter. 
More precisely, we shall restrict to the following situation:
\begin{enumerate}
\item the Lax operator is a $r \times r$ matrix polynomial of degree $n$ in the spectral parameter,
$L(\lambda)=A\lambda^n+M_1\lambda^{n-1}+\cdots+M_n$, with a
\emph{constant} leading term $A\in\gl(r)$; 
\item the constant matrix $A$ should commute only with linear combinations of
its powers (including $A^0\equiv\identity$). Equivalently, if
$A$ is diagonalisable, it should have distinct eigenvalues; more generally, the canonical Jordan
form of
$A$ should not contain Jordan blocks \emph{proportional to the identity} of dimension higher than
one. This property is generic on $\gl(r)$, but the requirement rules out some cases considered in
the literature \cite{Magri2}\cite{Magri2b}. A consequence of this requirement is that $A$ may be
nilpotent, with
$A^r=0$, but $A^k$ should not vanish for $k<r$.  
\item the variable matrices $M_i$ are generic matrices belonging to $\gl(r)$: we are not
considering restrictions to proper subalgebras, nor other types of reductions.
\end{enumerate}
This setting includes classical models such as the Euler--Poinsot top
(for $r=3$, $n=1$) and the Lagrange top (for $r=3$, $n=2$), in the sense which has already been
explained in the introduction: the classical models are properly embedded into larger systems,
but can be recovered by simple restriction (of a subset of the flows) on the appropriate invariant
submanifold (namely, the subalgebra of antisymmetric matrices). Other important examples such as
the Kovalewska top, or the Dubrovin--Novikov finite--dimensional reductions \cite{DubNov} of the
Gel'fand--Dickey soliton hierarchies, are strictly related to the systems that we are considering
but cannot be directly obtained by simple restriction \emph{``a posteriori''}. The
application of our framework to the periodic Toda lattice or to other models without a constant
leading term in the Lax matrix has not been investigated yet.

There are two possible approaches to the construction of multihamiltonian structures for Lax
equations of the type considered. Most authors regard them as dynamical systems on loop algebras
$\tilde{\gl}(r)\equiv {\gl}(r)((\lambda))$, and use the $R$--matrix technique to define compatible
Poisson brackets, which can be reduced to finite--dimensional quotient spaces identified with
the linear spaces of fixed--order polynomials in $\lambda$ \cite{ReySem1}. 

According to another
approach, one considers the direct sum of $n$ copies of the Lie algebra $\gl(r)$, defines a
suitable Lie algebra structure on this vector space (different from the direct product structure),
and an appropriate scalar product; in this way, one gets a natural Lie--Poisson bracket on
$\gl(r)^n$. The other Poisson structures are obtained by a \emph{deformation} procedure, i.e.~are
defined as Lie derivatives of the Lie--Poisson tensor along suitable vectorfields \cite{MagnMagri}.
In this approach, the dynamical variable is a $n$-ple of matrices $(M_1,\ldots,M_n)$, while the
fixed matrix
$A$ occurs in the definition of the deformation vectorfields which produce the Poisson structures;
the Lax matrix $L(\lambda)$ itself arises as a by--product, in connection with the
Hamilton equations with spectral parameter which naturally represent the trihamiltonian flows.
The two approaches are substantially equivalent for our purposes. We shall follow here the second
approach, but we stress that most of the definitions could be rephrased in the $R$--matrix
language. 

As anticipated in the introduction, we are not giving here all the proofs. The proof that the
tensors $\Pp$ and $\Q$ defined below are compatible Poisson tensors can be found, for instance, in
\cite{MagnMagri} or \cite{ReySem1}. The only proof which is
included concerns the fact that the characteristic determinant of the Lax matrix is the fundamental
common Casimir polynomial for our trihamiltonian structure.

\subsection{Affine Lie--Poisson pencils}

Let us consider the linear space \mbox{$\gl(r)^n \equiv \bigoplus^n\gl(r)$}; we shall denote
elements of this space by $\mathbf{M}\equiv(M_1,\ldots,M_n)$, with $M_i\in\gl(r)$. The scalar
product on $\gl(r)$ defined by (\ref{metrica}) is extended ``componentwise'' to $\gl(r)^n$:
\BE
(\mathbf{A},\mathbf{B}) = \sum_{i=1}^n\tr(A_i\cdot B_i).\label{metrican}
\EE
Using this scalar product, the gradient of a function $f:\gl(r)^n\rightarrow\Reali$ is again an
element of $\gl(r)^n$, that we denote by $(\nabla_1 f,\ldots,\nabla_n f)$. We define on
$\gl(r)^n$ a first Lie-Poisson structure $\Pn$ (depending on $M_i$ in a strictly linear way) by
setting 
\BE\label{Pn}
{\Pois{f}{g}}_{_{\Pn}} = \sum_{i=1}^n \Scal{\nabla_i g}{\sum_{k=i}^n 
\Lie{M_k}{\nabla_{k-i+1} f}}.
\EE
It is convenient to represent the Poisson tensor $\Pn$ as a
\emph{matrix of linear operators} acting on the column vector $(\nabla_1 f,\ldots,\nabla_n
f)$:
\BD\label{matPn}
\Pn = \left(
\begin{array}{ccccc}
\Lie{M_1}{\lcdot}     & \Lie{M_2}{\lcdot} & \cdots & 
\Lie{M_{n-1}}{\lcdot} & \Lie{M_n}{\lcdot} \\
\Lie{M_2}{\lcdot}     & \Lie{M_3}{\lcdot} & \cdots &   
\Lie{M_n}{\lcdot}   &        0         \\
\vdots           &      \vdots      &        &       \vdots         
&     \vdots      \\
\Lie{M_{n-1}}{\lcdot} & \Lie{M_n}{\lcdot} & \cdots &          
0           &        0         \\
\Lie{M_n}{\cdot}   &         0        & \cdots &          0           
&        0
\end{array}
\right)
\ED
Following \cite{MagnMagri}, from this first structure it is possible 
to obtain a \emph{sequence} of other
\mbox{$n$} \emph{affine Poisson structures} $P^{(n-1)},\ldots,P^{(0)}$, all mutually compatible,
by the iterative formula
\BE\label{alpp}
P^{(k-1)} = -\frac{1}{k+1} \mathcal{L}_X P^{(k)} \qquad k = n \ldots 1
\EE
where the components of the deformation vectors field $X$ are \emph{affine} functions, determined by a
the fixed matrix $A\in\gl(r)$:
\BD
X = \left({\tiny
\begin{array}{ccccc}
0   & \cdots & \cdots & \cdots &    0   \\
n-1  & \ddots &        &        & \vdots \\
0   & \ddots & \ddots &        & \vdots \\
\vdots & \ddots & \ddots & \ddots & \vdots \\
0   & \cdots &    0   &    1   &    0
\end{array}}
\right)
\left({\tiny
\begin{array}{c}
M_1 \\
\vdots \\
\vdots \\
\vdots \\
M_n \end{array}} \right) + \left({\tiny\begin{array}{c} n A \\[5pt]
0 \\
\vdots \\
\vdots \\[5pt]
0
\end{array}}
\right) \,.
\ED
The first pencil of the trihamiltonian structure that we shall consider is defined by
the tensors $P^{(1)}$ and $P^{(0)}$ of the sequence; their
expression is\goodbreak
\begin{eqnarray*}
{\Pois{f}{g}}_{_{P^{(1)}}} &=& \Scal{\nabla_1 g}{\Lie{\nabla_{n-1} f}{A}} +
\sum_{i=2}^{n-1} \Big({\nabla_i g},{\Lie{\nabla_{n-i} f}{A}+
\sum_{k=1}^{i-1} \Lie{\nabla_{n-i+k} f}{M_k}}\Big)\\
&&+\Scal{\nabla_n g}{\Lie{M_n}{\nabla_n f}} \\ \\
{\Pois{f}{g}}_{_{P^{(0)}}} &=& \Scal{\nabla_1 g}{\Lie{\nabla_n f}{A}} +
\sum_{i=2}^n \Big({\nabla_i g},{\Lie{\nabla_{n-i+1} f}{A}
+\sum_{k=1}^{i-1} \Lie{\nabla_{n-i+k+1} f}{M_k}}\Big).
\end{eqnarray*}
In matrix representation,
\begin{eqnarray}\label{pencil1}
P^{(1)} &=& \left(
\begin{array}{ccccc}
0 & \cdots & 0 & \Lie{\lcdot}{A} & 0 \\
0 & \cdots & \Lie{\lcdot}{A} & \Lie{\lcdot}{M_1} & 0 \\
\vdots & & & \vdots & 0 \\
0 & \Lie{\lcdot}{A} & \cdots & \Lie{\lcdot}{M_{n-3}} & 0 \\
\Lie{\lcdot}{A} & \Lie{\lcdot}{M_1} & \cdots & \Lie{\lcdot}{M_{n-2}} & 0 \\
0 & \cdots & \cdots & 0 & \Lie{M_n}{\lcdot}
\end{array}
\right) \nonumber\\ \\
P^{(0)} &=& \left(
\begin{array}{ccccc}
0 & \cdots & \cdots & 0 & \Lie{\lcdot}{A} \\
0 & \cdots & \cdots & \Lie{\lcdot}{A} & \Lie{\lcdot}{M_1} \\
\vdots & & & \vdots & \vdots \\
0 & \Lie{\lcdot}{A} & \cdots \cdots & & \Lie{\lcdot}{M_{n-2}} \\
\Lie{\lcdot}{A} & \Lie{\lcdot}{M_1} & \cdots \cdots & & \Lie{\lcdot}{M_{n-1}} 
\end{array}
\right)\nonumber
\end{eqnarray}

The Poisson pencil \mbox{$P^{(1)} - \lambda P^{(0)}$} is called the 
\emph{affine Lie-Poisson pencil} on $\gl(r)$. From the matrix representation given above, it
is easy to see that any function $f_\lambda=\sum h_i\lambda^i$ such that
\BE\label{caslp}\left\lbrace
\begin{array}{rccc}
\nabla_k f_{\lambda} - \lambda \nabla_{k+1} f_{\lambda} &=& 0\ ,&\qquad k=1,\ldots, n-1 \\
\Lie{A\lambda^n+M_1\lambda^{n-1}+\cdots+M_n}{\nabla_n f_{\lambda}} &=& 0 \ .&
\end{array}\right.
\EE
is a Casimir function of the affine Lie-Poisson pencil. This
observation leads to the definition of the Lax matrix
\mbox{$L(\lambda)=A\lambda^n+M_1\lambda^{n-1}+\cdots+M_n$}, and one immediately sees that the trace
of any power of $L(\lambda)$ fulfills (\ref{caslp}). Each Casimir function of
this type generate a bihamiltonian hierarchy according to the prescriptions (\ref{hamspec1}) and
(\ref{hamspec2}).  The corresponding Hamilton equations with spectral parameter are equivalent to
Lax equations for
$L(\lambda)$: the $k$-th flow of the hierarchy, using the same notation as in (\ref{hamspec1}),
is represented by
\BE\label{parlax}
\dot{L}(\lambda) = \left[{L(\lambda)},{\nabla_n f_\lambda^{(k)}}\right].
\EE

In the sequel, we revert to the notation used in the previous part of the paper, setting
$\Pp\equiv P^{(0)}$ and $\Q\equiv P^{(1)}$. However, on $\gl(r)^n$ the
third compatible structure $\R$, necessary to construct the appropriate second pencil, does
\emph{not} belong to the above--defined sequence of affine Lie--Poisson tensors: in particular, it
has nothing to do with the structure $P^{(2)}$, defined by (\ref{alpp}) for $n\ge 2$ (for this
reason we had to adopt a different notation). 

Our method to find the third Poisson tensor $\R$ still starts from the linear
Lie-Poisson structure (\ref{Pn}), but instead of applying a deformation vectorfield we exploit the
\emph{Lax-Nijenhuis equation}. This equation has been introduced in \cite{Magri4} in
connection with the following question: \emph{``Given a Lax equation, and a Poisson structure $P$
for which the traces of the powers of the Lax matrix are in involution, does it exists a second
compatible Poisson structure $Q$ such that the same constants of motion are iteratively linked in a
Magri--Lenard hierarchy?''} It turns out that, if such a second Poisson structure exists, the two
derivatives of the Lax matrix $L$ along the two vectorfields $Pdh$ and
$Qdh$ generated by any hamiltonian $h$ are linked by the following relation:
\BD
\mathcal{L}_{Qdh }L = \half\mathcal{L}_{Pdh}(L^2) + \Lie{L}{\alpha(dh)} ,
\ED
for some matrix $\alpha$ algebraically depending on differential of the hamiltonian $h$. In some
cases, this equation allows one to determine completely the second Poisson structure $Q$. This
may happen if the Lax matrix is not generic, and in particular if $L$ is assumed to have a fixed
degree with respect to some grading in the Lie algebra; for instance, when $L$ is tridiagonal
(Toda), or is a polynomial of fixed degree in a spectral parameter (the case we are dealing with).
Then, its square $L^2$ has usually a different degree, and the unknown element $\alpha$ occurring
in the r.h.s.~become determined by a \emph{compatibility} requirement, i.e.~its commutator with $L$
should cancel exactly the terms of higher degree in the derivative of $L^2$. We refer the reader to
\cite{Magri4} for a complete presentation of the method.

We omit the details of the computation in our case: a key point is that one should
plug in the Lax--Nijenhuis a slightly modified Lax matrix polynomial, namely the ``convoluted''
polynomial $L^*(\lambda) = A + M_1\lambda + \cdots + M_n\lambda^n$. Then, 
starting from the Poisson tensor $\Pp$, one finds the
following new Poisson bracket: 
\begin{eqnarray}\label{R}
{\Pois{f}{g}}_{\R} & = & \sum_{i=1}^n \Big(\nabla_i g,\sum_{k=i}^n 
\Triplo{M_k}{\nabla_{k-i+1} f}{A}\Big) +
\\
&& \sum_{i=2}^n \Big(\nabla_i g,\sum_{k=i}^n \sum_{l=1}^{i-1} 
\Triplo{M_k}{\nabla_{k+l-i+1} f}{M_l}\Big) \,.\nonumber
\end{eqnarray}

\noindent For $n$ generic, writing down the representation of this Poisson tensor as a matrix of
linear operators, analogous to
(\ref{pencil1}), would be rather cumbersome. The reader can easily figure out the general form
from the representations of the Poisson tensors respectively corresponding to $n=1,2,3$:
\begin{eqnarray*}
\mathrm{for}\ n=1, \qquad &\R =& \Triplo{M}{(\cdot)}{A}\\[20 pt]
\mathrm{for}\ n=2, \qquad &\R =& \left(
\begin{array}{cc}
\Triplo{M_1}{(\lcdot)}{A} & \Triplo{M_2}{(\lcdot)}{A} \\[10pt]
\Triplo{M_2}{(\lcdot)}{A} & \Triplo{M_2}{(\lcdot)}{M_1} 
\end{array}
\right)\\[20 pt]
\mathrm{for}\ n=3, \qquad &\R =& \left({\tiny
\begin{array}{ccc} \Triplo{M_1}{(\lcdot)}{A} & 
\Triplo{M_2}{(\lcdot)}{A}  & \Triplo{M_3}{(\lcdot)}{A} \\[15pt]
\Triplo{M_2}{(\lcdot)}{A} & \begin{array}{c} \Triplo{M_3}{(\lcdot)}{A} 
+ \\
\Triplo{M_2}{(\lcdot)}{M_1}
\end{array}
& \Triplo{M_3}{(\lcdot)}{M_1} \\[15pt]
\Triplo{M_3}{(\lcdot)}{A} & \Triplo{M_3}{(\lcdot)}{M_1} & 
\Triplo{M_3}{(\lcdot)}{M_2}
\end{array}}
\right)
\end{eqnarray*}
For $n=1$ one recovers the Poisson structure of Morosi and Pizzocchero \cite{MorPizz} already
described in the Introduction. The quadratic Poisson structures for $n>1$ have never been presented
in the previous literature, to our knowledge. In the $R$--matrix language, they can be obtained
by a suitable modification of the so--called Sklyanin bracket \cite{Sem}.

\subsection{Fundamental Casimir polynomial}

We have anticipated in the introduction that the characteristic determinant of the Lax matrix
$L(\lambda)$ provides a complete common Casimir polynomial for the two pencils 
\mbox{$\Q - \lambda \Pp$} and \mbox{$\R - \mu \Pp$}. We shall now prove this statement.

We already know that the traces of the powers of the Lax matrix are Casimir functions for the
first pencil, as they fulfill (\ref{caslp}). The same holds for the coefficients of each power of
$\mu$ in the characteristic polynomial $f_{\lambda\mu}=\det(L(\lambda) - \mu \identity)$, since
these coefficients are functionally dependent on the traces of the powers of $L(\lambda)$.
Then, what remains to check is that the polynomials in $\mu$ which occur as coefficients of each
power of $\lambda$ in $f_{\lambda\mu}$ are Casimir functions for the second pencil.

In the following computation, we denote the components of the gradient of $f_{\lambda\mu}$ by
\mbox{$ V_i = \nabla_i f_{\lambda,\mu} $} (notice that each matrix $V_i$ still depends on both
$\lambda$ and $\mu$), and we systematically use the fact that
\BE\label{conditio}
V_k = \lambda V_{k+1}, \qquad \mathrm{i.e.} \qquad V_k = \lambda^{n-k}V_n.
\EE
\noindent The condition $(\R - \mu \Pp)\, df_{\lambda\mu}=0$ translates into the following system of
$n$ equations:
\BD
\left\lbrace
\begin{array}{lr}
\Lie{\mu V_n}{A} - \sum_{k=1}^n(\Triplo{M_k}{V_k}{A}) = 0 \\[10pt]
\Lie{\mu V_{n-i+1}}{A} +
\sum_{k=1}^{i-1}\Lie{\mu V_{n-i+k+1}}{M_k}\\[5pt] -
\sum_{k=i}^n(\Triplo{M_k}{V_{k-i+1}}{A})\\[5pt] - 
\sum_{k=i}^n\sum_{l=1}^{i-1}(\Triplo{M_k}{V_{k+l-i+1}}{M_l}) = 0 &
\mathrm{for}\ i=2,\ldots,n 
\end{array}\right.
\ED
which, with some algebraic manipulations taking account of 
(\ref{conditio}) \emph{and} of some of the equalities following from $(\Q - \lambda \Pp)\,
df_{\lambda\mu}=0$, can be shown to be equivalent to
\BE\label{Cas2}
\left(L(\lambda) - \mu \identity\right) V_n M_i = M_i V_n \left(L(\lambda) - \mu \identity\right)
\quad 
\mathrm{for\ all}\ i=1\ldots,n
\EE
Now, for the characteristic polynomial $f_{\lambda\mu}=\det(L(\lambda) - \mu \identity)$
one has
\BD
V_n = \nabla_n f_{\lambda\mu} = f_{\lambda\mu}\cdot\left(L(\lambda) - \mu
\identity\right)^{-1},
\ED
therefore equation (\ref{Cas2}) is straigthforwardly satisfied.

Under the hypotheses listed at the beginning of this section (the phase space is the full space
$\gl(r)^n$, without constraints, and $A$ is suitably generic), the non constant
coefficients of the characteristic equation are functionally independent and define exactly
$\half{nr(r+1)}$ hamiltonians. Fig.~1 displays the corresponding ``fundamental molecule''. One sees
directly from the diagram that, for each one of the Poisson tensors $\Pp$, $\Q$ and
$\R$, the set of hamiltonians $h_i^j$ includes exactly $nr$ Casimir functions, therefore the rank
of the Poisson tensor cannot exceed $nr^2-nr = nr(r-1)$; on the other hand, the remaining
$\half{nr(r-1)}$ hamiltonians are in mutual involution, so the rank of the Poisson tensor cannot be
less than $nr(r-1)$. Therefore, the fundamental Casimir polynomial
$f_{\lambda\mu}$ is complete. If $A$ does not fulfill our requirements, what happens is that (i)
the Poisson structures $\Q$ and $\R$ have a larger kernel, so there are other independent Casimir
functions not occurring in the characteristic determinant; (ii) some coefficients of the
characteristic determinant vanish identically, and by consequence (iii) one cannot find properly
normalised transversal vectorfields using the recipe presented below. 

\subsection{Transversal vectorfields and separation coordinates}

The next step towards the construction of separation coordinates consists in finding a set of
$\Pp$--transversal vectorfields, normalized on the $nr$ Casimir functions which one gets from the
fundamental Casimir polynomial. We state without proof the general recipe:
\begin{enumerate}
\item Choose a matrix $W_1\in\gl(r)$ of rank one such that:
\begin{eqnarray}\label{wsubs}
\tr(W_1) &=& 0 \nonumber\\
\tr(W_1 A) &=& 0 \nonumber\\
&\vdots \\
\tr(W_1 A^{r-2}) &=& 0 \nonumber\\
\tr(W_1 A^{r-1}) &=& (-1)^{r-1};\nonumber
\end{eqnarray}
this condition makes sense because we know that $A^k\not= 0$ for $k<r$ (the $r$ linear
equations above do not determine uniquely $W_1$, but any
such $W_1$ will work).

\item Compute the \emph{adjoint} of the characteristic matrix
$L(\lambda)-\mu\identity$, that we shall denote by $L^\dag(\lambda,\mu)$; by definition, 
$\big(L(\lambda)-\mu\identity\big)\cdot L^\dag(\lambda,\mu)=f_{\lambda\mu}\identity$. The
entries of $L^\dag(\lambda,\mu)$ are the cofactors of $L(\lambda)-\mu\identity$, therefore
$L^\dag(\lambda,\mu)$ is polynomial of order $n(r-1)$ in $\lambda$ and of order $(r-1)$ in $\mu$.

\item Let $W_{\lambda}$ be the matrix polynomial
of the form
\BE\label{antilax}
W_{\lambda} = W_1\lambda^{n-1} + W_2 \lambda^{n-2} + \cdots + W_{n}.
\EE
with
\begin{eqnarray}
W_2 &=& W_1\,\big(u_{0,2} \identity + u_{1,2} A + \cdots + u_{r-2,2} A^{r-2}\big)\nonumber\\
&\vdots& \\
W_n &=& W_1\,\big(u_{0,n} \identity + u_{1,n} A + \cdots + u_{r-2,n} A^{r-2}\big)\nonumber.
\end{eqnarray}
The $(r-1)\times (n-1)$ coefficients $u_{i,j}$ are scalar functions on $\gl(r)^n$,
which should be determined by the following condition: 
\emph{take the polynomial
$\tr\big(L^\dag(\lambda,\mu)\, W_{\lambda}\big)$; for each power of $\mu$ separately, the $n$
highest coefficients in $\lambda$ should vanish,} except for the coefficient of
$\lambda^{n(r-1)+(n-1)}$, which is always equal to one because of (\ref{wsubs}). 
Namely, the terms which should be canceled with the appropriate choice of $u_{i,j}$ are the
following: 
\BD
\begin{array}{crcl}
\lambda^{k}, & n(r-1)&\le k \le& n(r-1)+(n-2), \\
\mu\lambda^{k},& (n-1)(r-1)&\le k \le& (n-1)(r-1)+(n-1),\\
\vdots \\
\mu^{n-1}\lambda^{k}, & 0&\le k \le& (n-1) 
\end{array}
\ED 
This gives a linear system of
equations for the coefficients $u_{i,j}$. The system is triangular and can always be solved.

\item Denoting a tangent vector on $\gl(r)^n$ by $\mathbf{v} = [\dot M_1,\ldots,\dot M_n]$,
introduce the
$n$ vectorfields
\begin{eqnarray}\label{icampi}
\mathbf{v}^{0}_1 &=& [0,\cdots,0,0,W_1] \nonumber\\
\mathbf{v}^{0}_2 &=& [0,\cdots,0,W_1,W_2] \nonumber\\
&\vdots& \\
\mathbf{v}^{0}_n &=& [W_1,\cdots,W_{n-1},W_{n}] \nonumber
\end{eqnarray}

\item Take for $k=1,\ldots,r-1$ the product of $W_{\lambda}$ with the
$k$--th power of the Lax matrix $L(\lambda)$:
\BE
W^{(k)}_{\lambda} = \frac{W_\lambda\,L^k}{\lambda^{nk}}.
\EE
Consider the highest $n$ terms in the expansion
$W^{(k)}_{\lambda} = W^{(k)}_1\lambda^{n-1} + W^{(k)}_2 \lambda^{n-2} + \ldots +
W^{(k)}_{n}\lambda^{0}+\ldots$ (in particular, one has
$W^{(k)}_1
\equiv W_1\,A^k$): other
$n(r-1)$ vectorfields
$\mathbf{v}^{k}_j$, for $i=1\ldots,n\,$ and
$\,k=1,\ldots,(r-1)$, are defined analogously to (\ref{icampi}): 
\begin{eqnarray}\label{icampi2}
\mathbf{v}^{k}_1 &=& [0,\cdots,0,0,W^{(k)}_1] \nonumber\\
\mathbf{v}^{k}_2 &=& [0,\cdots,0,W^{(k)}_1,W^{(k)}_2] \nonumber\\
&\vdots& \\
\mathbf{v}^{k}_n &=& [W^{(k)}_1,\cdots,W^{(k)}_{n-1},W^{(k)}_{n}] \nonumber
\end{eqnarray}

\end{enumerate}
For the $nr$ vectorfields constructed in this way, the following holds:
\begin{Prop}{}
The vectorfields $\mathbf{v}^{k}_j$ are symmetries of $\Pp$ and fulfill the normalization condition
\BE
\mathbf{v}^{k}_j(h^l_{m+(r-k-1)n})=\delta^{kl}\delta_{jm}\quad \mbox{for }
k,l=0,\ldots(r-1)\mbox{ and } j,m=1,\ldots,n;
\EE
moreover, for all $k$, $j$, $l$ and $m$ in the given ranges,
\BE\label{secder}
\mathbf{v}^{k}_j\left(\mathbf{v}^{l}_m\left(f_{\mu\nu}\right)\right)=0\quad and\quad
[\mathbf{v}^{k}_j,\mathbf{v}^{l}_m]=0
\EE
\end{Prop}

To ensure that the $\Pp$--transversal vectorfields $\mathbf{v}^{k}_j$ provide a set of (separation)
Darboux--Nijenhuis coordinates, one should also check conditions (2) and (3) of Prop.(3.7).
However, this verification may be postponed: let us tentatively assume that these
conditions are satisfied. Then, one could find separation coordinates in which the Poisson tensors
$\Pp$, $\Q$ and $\R$ assume the ``canonical'' form described in section (3.3); in particular, the
affine Lie--Poisson tensor $\Pp$ becomes canonical in the usual sense. 

In separation coordinates, we already know the explicit form of the $nr$ polynomials
$S_{\alpha}(\lambda,\mu)$; the latter should coincide, up to the change of coordinates, with the
polynomials
$S^{k}_j(\lambda,\mu)\equiv\mathbf{v}^{k}_j\left(f_{\lambda\mu}\right)$ that one can also compute in
terms of the original variables of the Lax equation. Hence, one should recover the separation
coordinates by taking the common roots of the latter polynomials. 
Actually, it is sufficient to compute the common roots of the polynomials 
$\mathbf{v}^{k}_1\left(f_{\mu\nu}\right)$. The vectorfields $\mathbf{v}^{k}_1$ are \emph{constant}:
namely, we have seen that they are defined by $\dot{M}_i=0$ for $i=1,\ldots,(n-1)$ and
$\dot{M}_n=W_1\,A^k$. The corresponding polynomials $S^{k}_1(\lambda,\mu)$ are the derivatives
of the characteristic determinant $f_{\lambda\mu}$ along constant vectors fields, and therefore
are nothing but \emph{linear combinations of cofactors of the characteristic matrix}
$L(\lambda)-\mu\identity$, with constant coefficients determined by the matrices $W_1$ and $A$.
Therefore, \emph{we recover Sklyanin's recipe, with a definite prescription of the
normalization to be used}\footnote{To be honest, exactly because Sklyanin's prescription does not
fix the normalisation of the Baker--Akhiezer function, to prove the equivalence of the two
procedures one cannot rely on the simple comparison on the final results, but
it would be necessary to compare each object involved in the two constructions, what
we have not done yet.} 

If one were able to compute $\half nr(r-1)$ pairs of common roots $(\lambda_i,\mu_i)$ of the
polynomials $S^{k}_1(\lambda,\mu)$, then one would only have to check \emph{a posteriori} that the
new variables $(\lambda_i,\mu_i)$ are canonical for the Poisson structure $\Pp$. Since the
polynomials $S^{k}_1(\lambda,\mu)$ are independent linear combinations of cofactors of the
characteristic matrix, $(\lambda_i,\mu_i)$ are also roots of the fundamental Casimir polynomial,
and this would
be enough to say that all the hamiltonians become separable in the new coordinates. In this
case there is no ambiguity on the constant part of $f_{\lambda\mu}$ itself, which turn out to depend
only on the choice of $A$. Notice that in the new coordinates on $\gl(r)^n$, given by
the roots $(\lambda_i,\mu_i)$ and by the $nr$ Casimir functions of
$\Pp$, the vectorfields $\mathbf{v}^{k}_j$ automatically become coordinate vectorfields, due to
the normalization and to the property (\ref{secder}). However, as we have stressed in the
introduction, finding explicitly the roots of the polynomials
$S^{k}_1(\lambda,\mu)$ is impossible in general. 

A more effective method is the following one: since we know the expression of the polynomials
$S^{k}_j(\lambda,\mu)$ in \emph{both} coordinate systems, we can simply equate the corresponding
coefficients for each polynomial to get a set of algebraic equations linking the two coordinate
systems. Luckily enough, while the coefficients of $S^{k}_j(\lambda,\mu)$ are in general rather
complicated rational functions of the separation coordinates, it is easy to see that a number
of coefficients are \emph{linear} functions of the entries of the Lax matrix $L(\lambda)$.
It turns out that one can provide in this way a complete set of equations relating the two
coordinate systems, which either are all linear (in the lower dimensional cases) or can be reduced
to linear equations. In this way one can effectively compute the \emph{inverse}
coordinate transformation, i.e.~express the original variables as functions of the separation
coordinates. The mapping cannot be explicitly inverted in general, but it is sufficient for some
purposes, for instance to check that the Poisson tensors transform as expected. 

Let us present a concrete example. We have already displayed in sect.(3.3) the form of the
polynomials $S_{\alpha}$ for the system associated to the Lie algebra $\gl(3)$. Let us now compute
the same polynomials for the Lax matrix $A\lambda+M$, with $M\in\gl(3)$ and $A$ diagonal with
distinct eigenvalues $(\auno,\adue,\atre)$ (this case corresponds to the
generalised Euler--Poinsot rigid body, discussed in the introduction). We choose a set of
\emph{orthonormal} coordinates $(x_1,\ldots,x_9)$ in $\gl(3)$, setting
\BE
M = \left (\begin {array}{ccc}
x_7 & \frac{1}{\sqrt{2}}(x_1+x_4) & \frac{1}{\sqrt{2}}(x_2+x_5) \\
\frac{1}{\sqrt{2}}(x_4-x_1) & x_8 & \frac{1}{\sqrt{2}}(x_3+x_6)\\
\frac{1}{\sqrt{2}}(x_5-x_2) & \frac{1}{\sqrt{2}}(x_6-x_3) & x_9
\end {array}
\right )
\EE
The three Casimir functions for $\Pp$ (see (\ref{cry1})) are
\begin{eqnarray}\label{lec}
h^0_2 &=& \adue \atre x_7 + \auno  \atre x_8 +  \auno \adue x_9 \nonumber\\
h^1_1 &=& -(\adue+\atre)x_7 -(\auno+\atre)x_8 -(\auno+\adue)x_9   \\
h^2_0 &=& x_7 + x_8 + x_9  \nonumber
\end{eqnarray}
Hence, there is a linear correspondence between the coordinates $(x_7,x_8,x_9)$ and the
coordinates $(c_1,c_2,c_3)$. Choosing $W_1$ to have all the rows equal to each other, following
step (1) above one finds
\BE
W_1 = \left(\begin {array}{ccc}
\frac{1}{(\auno-\adue) (\auno-\atre)} & \frac{1}{(\atre-\adue) (\auno-\adue)} & \frac{1}{(\adue-\atre) (\auno-\atre)} \\
\frac{1}{(\auno-\adue) (\auno-\atre)} & \frac{1}{(\atre-\adue) (\auno-\adue)} & \frac{1}{(\adue-\atre) (\auno-\atre)} \\
\frac{1}{(\auno-\adue) (\auno-\atre)} & \frac{1}{(\atre-\adue) (\auno-\adue)} & \frac{1}{(\adue-\atre) (\auno-\atre)} \\
\end {array}
\right )
\EE
There are only three constant transversal vectorfields in this case, with
$\dot{M}=W_1$, $\dot{M}=W_1\,A$ and $\dot{M}=W_1\,A^2$ respectively.
Taking the derivatives of $f_{\lambda\mu}=\det(A\lambda+M-\mu\identity)$ along these three
vectorfields one easily computes the three polynomials $S_1(\lambda,\mu)$, $S_2(\lambda,\mu)$ and
$S_3(\lambda,\mu)$. Let us write down only the coefficients which are useful for our
purpose:
\begin{eqnarray*}
S_1 &=&
\lambda^2 + \frac{1}{\sqrt{2}}\Big[-(\auno-2\atre+\adue)x_1+(\auno-2\adue+\atre)x_2 +
(2\auno-\adue-\atre)x_3 \\ &&  + \sqrt{2}(\atre-\adue)x_7 + \sqrt{2}(\auno-\atre)x_8 +
\sqrt{2}(\adue-\auno)x_9 \\ &&   + (\auno-\adue)x_4 -(\auno-\atre)x_5+(\adue-\atre)x_6\Big]
\left[(\atre-\adue) (\auno-\adue) (\auno-\atre) \right]^{-1} \mu \\ &&
+\frac{1}{\sqrt{2}}\Big[\atre(\auno-2\atre+\adue) x_1-\adue(\auno-2\adue+\atre)x_2 -
\auno(2\auno-\adue-\atre)x_3 \\ &&
+\sqrt{2}(\atre-\adue)(-\atre+\auno-\adue)x_7 - \sqrt{2}(\auno-\atre)(\auno-\adue+\atre)x_8 
\\ &&+\sqrt{2}(\auno-\adue)(\auno-\atre+\adue)x_9
 + \auno(\atre-\adue)x_6 
 - \atre(\auno-\adue)x_4 \\ && +\adue(\auno-\atre)x_5
\Big] \left[(\atre-\adue) (\auno-\adue) (\auno-\atre) \right]^{-1} \lambda + \ldots \\
\\
S_2 &=&
\lambda\mu + \frac{1}{\sqrt{2}}\Big[(2\auno\atre-\adue\atre-\auno\adue)x_2 + (\auno\atre-2\adue\atre+\auno\adue)x_3 \\
&& +(-2\auno\adue+\adue\atre+\auno\atre)x_1 + \sqrt{2}(\auno-\atre)\adue x_8 -
\sqrt{2}(\auno-\adue)\atre x_9 \\ && -\auno(-\adue+\atre)x_6 + \sqrt{2}\auno(-\adue+\atre)x_7 +
\atre(\auno-\adue)x_4 \\ && -\adue(\auno-\atre)x_5 \Big] \left[(\atre-\adue) (\auno-\adue)
(\auno-\atre) \right]^{-1} \mu \\ &&
+\frac{1}{\sqrt{2}}\Big[-\adue(2\auno\atre-\adue\atre-\auno\adue)x_2 \\ &&
-\auno(\auno\atre-2\adue\atre+\auno\adue)x_3 - \atre(-2\auno\adue+\adue\atre+\auno\atre)x_1 \\ &&
-\sqrt{2}\auno\atre(\auno-\atre)x_8 + \sqrt{2}\auno\adue(\auno-\adue)x_9 + \auno^2(-\adue+\atre)x_6
\\ && -\sqrt{2}\atre\adue(-\adue+\atre)x_7 - \atre^2(\auno-\adue)x_4 \\ && +\adue^2(\auno-\atre)x_5
\Big] \left[(\atre-\adue) (\auno-\adue) (\auno-\atre) \right]^{-1} \lambda +\ldots \\
\\
S_3 &=&
\lambda^2 + \frac{1}{\sqrt{2}}\Big[(\atre^2\auno-\adue\atre^2-\auno^2\adue +\auno^2\atre)x_2 
 +(\atre^2\auno-\adue\atre^2+\adue^2\auno-\adue^2\atre)x_3 \\
&& +(-\adue^2\auno+\adue^2\atre-\auno^2\adue+\auno^2\atre)x_1 
 -\sqrt{2}(-\auno\adue+\auno\atre-\adue\atre)(\auno-\atre)x_8 \\
&& -\sqrt{2}(-\auno\adue+\auno\atre+\adue\atre)(\auno-\adue)x_9 
 -(\auno\atre-\adue\atre+\auno\adue)(-\adue+\atre)x_6 \\
&& +\sqrt{2}(\auno\atre-\adue\atre+\auno\adue)(-\adue+\atre)x_7 
 +(-\auno\adue+\auno\atre+\adue\atre)(\auno-\adue)x_4 \\
&& +(-\auno\adue+\auno\atre-\adue\atre)(\auno-\atre)x_5\Big]\cdot\\
&&\hskip3cm\cdot \left[(\atre-\adue) (\auno-\adue) (\auno-\atre) \right]^{-1} \mu\\
&& +\frac{1}{\sqrt{2}}\Big[-\atre(-\adue^2\auno+\adue^2\atre-\auno^2\adue+\auno^2\atre)x_1 
-\sqrt{2}\auno\atre\adue(-\adue+\atre)x_7\\
&& -\adue(\atre^2\auno-\adue\atre^2-\auno^2\adue+\auno^2\atre)x_2
-\sqrt{2}\auno\adue\atre(\auno-\atre)x_8\\ 
&& -\auno(\atre^2\auno-\adue\atre^2+\adue^2\auno-\adue^2\atre)x_3 
+\sqrt{2}\auno\adue\atre(\auno-\adue)x_9\\
&&   - (-\auno\adue+\auno\atre+\adue\atre)(\auno-\adue)\atre
x_4 \\ && -(-\auno\adue+\auno\atre-\adue\atre)(\auno-\atre)\adue x_5\Big]\cdot \\
&&+(\auno\atre-\adue\atre+\auno\adue)(-\adue+\atre)\auno x_6 \\
&&\hskip3cm\cdot \left[(\atre-\adue) (\auno-\adue) (\auno-\atre) \right]^{-1} \lambda +\ldots
\end{eqnarray*}
Therefore, there are exactly six coefficients which are linear in the coordinates $x_i$.
One can check that they are independent and also independent from the other three equations already
found from (\ref{lec}): the determinant of the full $9\times 9$ linear system is identically equal
to one. By equating the polynomials above to the corresponding expressions listed
at the end of sect.(3.3), one finds the coordinate transformation. On the other hand, all three
polynomials contain quadratic terms in $\lambda$ and $\mu$, and to find the common roots of two of
them one should solve an equation of order four in $\lambda$ (or in
$\mu$), which is possible in principle but gives a result which is of little practical use.

The generalization of the Lagrange top can be treated in the same way. The phase space is
$\gl(3)^2$, and as above we use as
coordinates suitable orthonormal linear combinations of the entries of $M_1$ and $M_2$, that we
denote by $x_i$, $i=1,\ldots,18$.  The constant matrix for this case is chosen to be
$
A=\tiny{\pmatrix{0 & 1 & 0 \cr
-1 & 0 & 0\cr
0 & 0 & 0}}.
$ 
The Casimir functions for $\Pp$ occurring in the characteristic determinant
$f_{\lambda\mu}=\det\left(A\,\lambda^2+M_1\lambda+M_2\right)$ are $h^0_4$, $h^0_5$,
$h^1_2$,
$h^1_3$,
$h^2_0$ and
$h^2_1$; they are linear except for $h^0_4$ and $h^1_2$ which are quadratic. There are 6 transversal vectorfields, three of which are
constant. Each of the three corresponding polynomials,
$S^0_4(\lambda,\mu)$, $S^1_2(\lambda,\mu)$ and $S^2_0(\lambda,\mu)$, has exactly three coefficients
which are linear in the coordinates $x_i$, namely the coefficients of $\mu$, $\lambda\mu$ and
$\lambda^3$. Then, one has a total of 13 independent linear equations which can be solved for the
13 coordinates $x_1,\ldots,x_9,x_{15},\ldots,x_{18}$. The coefficients of $\lambda$ and $\lambda^2$
in the polynomials $S^0_4$, $S^1_2$ and $S^2_0$, are linear in the remaining five coordinates
$x_{10},\ldots,x_{14}$, which can thus be computed as well. On the contrary, the direct procedure
\emph{\`a la Sklyanin}, i.e.~computing the common  roots of the bivariate polynomials $S_\alpha$,
is not viable due to the order of the polynomials themselves.

\subsection*{Acknowledgements}
We are much indebted to Franco Magri, Gregorio Falqui, Marco Pedroni
and Sergio Benenti for several helpful discussions and exchanges of
information on work in progress, and to the referee for useful
comments. This work is supported by the national MURST research
project "Geometry of Integrable Systems". 
\vfill\eject

\newsavebox{\molecola}
\savebox{\molecola}(650,300){
\Scatolona{0}{70}{$0$}
\put(15,20){\NucAlt{$h^0_0$}}
\Scatolona{30}{0}{$0$}
\Scatolona{57.5}{70}{$X^0_0$}
\put(62.5,95){\usebox{\Rdir}}
\put(70,20){\NucAlt{$h^0_1$}}
\Scatolona{85}{0}{$0$}
\Scatolona{112.5}{70}{$X^0_1$}
\put(117.5,95){\usebox{\Rdir}}
\Scatolona{40}{195}{$0$}
\put(55,145){\NucAlt{$h^{n-2}_0$}}
\Scatolona{70}{125}{$\cdots$}
\Scatolona{97.5}{195}{$X^{n-2}_0$}
\put(110,145){\NucAlt{$h^{n-2}_1$}}
\Scatolona{125}{125}{$\cdots$}
\Scatolona{152.5}{195}{$X^{n-2}_1$}
\Scatolona{67.5}{270}{$0$}
\put(82.5,220){\NucAlt{$h^{n-1}_0$}}
\Scatolona{125}{270}{$0$}
\put(137.5,220){\NucAlt{$h^{n-1}_1$}}
\Scatolona{180}{270}{$0$}
\put(192.5,260){\usebox{\Qdir}}
\Scatolona{207.5}{235}{$\cdots$}
\put(232.5,260){\usebox{\Pdir}}
\Scatolona{235}{270}{$0$}
\put(220,145){\NucAlt{$h^{n-2}_{r-1}$}}
\Scatolona{235}{125}{$\cdots$}
\Scatolona{262.5}{195}{$X^{n-2}_{r-1}$}
\put(275,145){\NucAlt{$h^{n-2}_r$}}
\Scatolona{290}{125}{$\cdots$}
\put(247.5,220){\NucAlt{$h^{n-1}_{r-1}$}}
\Scatolona{292.5}{270}{$0$}
\Scatolona{317.5}{195}{$0$}
\put(330,185){\usebox{\Qdir}}
\Scatolona{345}{160}{$\cdots$}
\put(370,185){\usebox{\Pdir}}
\Scatolona{372.5}{195}{$0$}
\put(385,145){\NucAlt{$h^{n-2}_{2r-1}$}}
\Scatolona{400}{125}{$\cdots$}
\Scatolona{430}{195}{$0$}
\put(400,20){\NucAlt{$h^0_{(n-2)r-1}$}}
\Scatolona{415}{0}{$0$}
\Scatolona{442.5}{70}{$X^0_{(n-2)r-1}$}
\put(447.5,95){\usebox{\Rdir}}
\put(455,20){\NucAlt{$h^0_{(n-2)r}$}}
\Scatolona{470}{0}{$0$}
\Scatolona{497.5}{70}{$0$}
\put(510,60){\usebox{\Qdir}}
\Scatolona{525}{35}{$\cdots$}
\put(550,60){\usebox{\Pdir}}
\Scatolona{552.5}{70}{$0$}
\put(565,20){\NucAlt{$h^0_{(n-1)(r-1)}$}}
\Scatolona{580}{0}{$0$}
\Scatolona{612.5}{70}{$0$}
}

\rotatebox{270}{
\begin{picture}(650,300)
\put(-70,50){
\framebox(650,300){\put(-325,0){\usebox{\molecola}}}
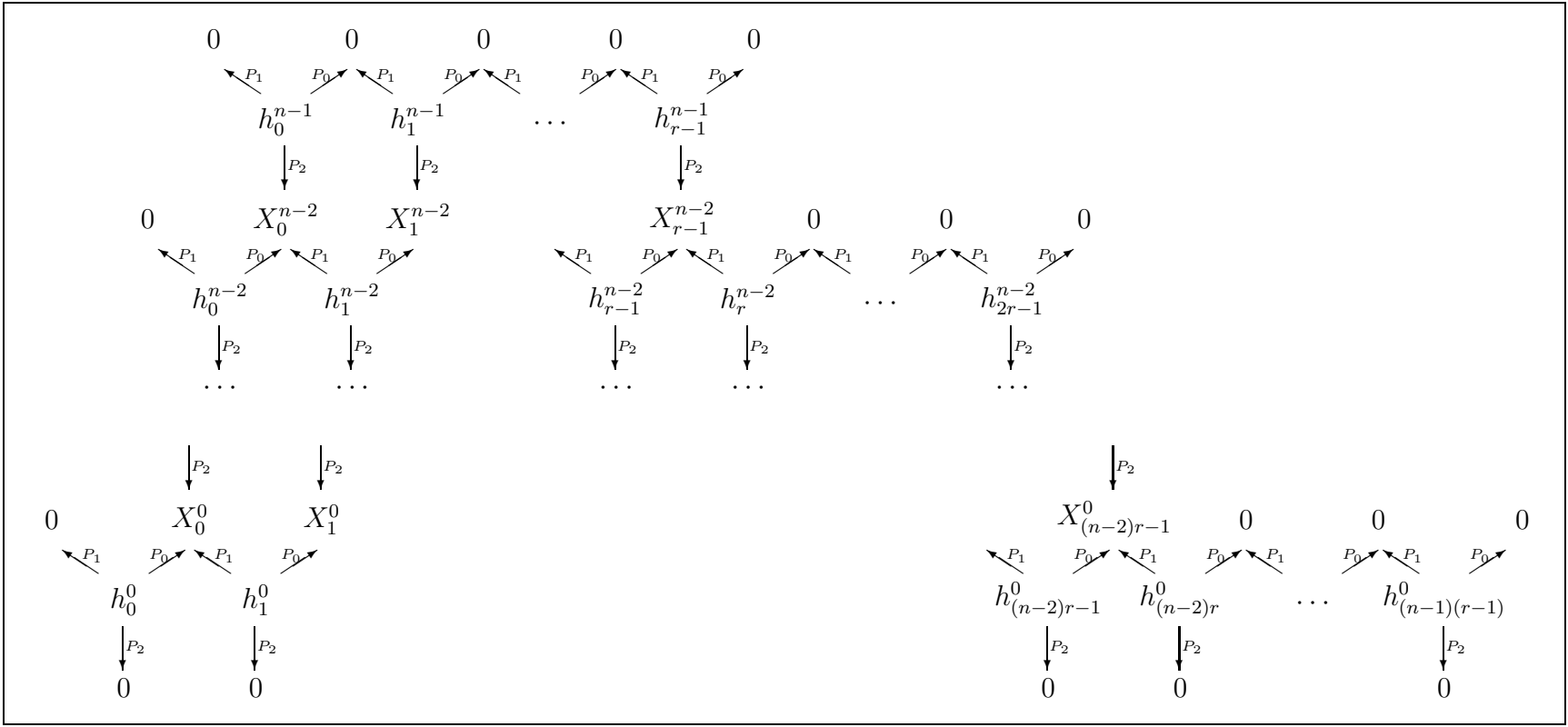
\put(-450,-50){\makebox{Figure 1: Fundamental molecule for $\Gotico{gl}(r)^n$}}
}
\end{picture}
}
\vfill\eject

\end{document}